\documentclass[]{jair}

\setcopyright{cc}
\copyrightyear{2025}
\acmDOI{10.1613/jair.1.17997}

\JAIRAE{Haris Aziz}
\JAIRTrack{} 
\acmVolume{83}
\acmArticle{15}
\acmMonth{7}
\acmYear{2025}

\RequirePackage[
  datamodel=acmdatamodel,
  style=acmnumeric,
  backend=biber,
  giveninits=true
  ]{biblatex}

\usepackage{graphicx}
\usepackage{microtype}
\usepackage{subfigure}
\usepackage{subcaption}
\usepackage{amsmath}
\usepackage{mathtools}
\usepackage{amsthm}
\usepackage{algorithm}
\usepackage{algorithmic}
\usepackage[capitalize,noabbrev]{cleveref}
\usepackage{url}
\usepackage{booktabs}
\usepackage{amsfonts}
\usepackage{nicefrac}
\usepackage{microtype}
\usepackage{xcolor}
\usepackage{dsfont}
\usepackage{float}
\usepackage{tikz}
\usetikzlibrary{shapes, positioning, arrows}

\DeclareMathOperator*{\R}{\mathbb{R}}
\DeclareMathOperator*{\N}{\mathbb{N}}

\DeclareMathOperator*{\argmax}{argmax}
\DeclareMathOperator*{\argmin}{argmin}
\DeclarePairedDelimiterX\set[1]\lbrace\rbrace{#1}
\def\lan{\lambda n}
\newcommand{\sumi}{\sum_{i=1}^{n}}
\newcommand{\sumj}{\sum_{j=1}^{n}}
\newcommand{\norm}[1]{\left\lVert#1\right\rVert}
\newtheorem{theorem}{Theorem}
\newtheorem{definition}{Definition}
\newtheorem{lemma}{Lemma}

\newtheorem{observation}{Observation}
\newtheorem{corollary}{Corollary}
\newtheorem{example}{Example}
\newtheorem{remark}{Remark}

\begin{document}

\title[The Search for Stability]{The Search for Stability: Learning Dynamics of Strategic Publishers with Initial Documents}

\author{Omer Madmon}
\authornote{Equal Contribution.}
\email{omermadmon@campus.technion.ac.il}
\affiliation{%
  \institution{Technion - Israel Institute of Technology}
  \city{Haifa}
  \country{Israel}
}

\author{Idan Pipano}
\authornote{Equal Contribution.}
\email{idan.pipano@campus.technion.ac.il}
\affiliation{%
  \institution{Technion - Israel Institute of Technology}
  \city{Haifa}
  \country{Israel}
}

\author{Itamar Reinman}
\authornote{Equal Contribution. Corresponding Author.}
\email{itamarr@campus.technion.ac.il}
\affiliation{%
  \institution{Technion - Israel Institute of Technology}
  \city{Haifa}
  \country{Israel}
}

\author{Moshe Tennenholtz}
\authornote{Equal Contribution.}
\email{moshet@technion.ac.il}
\affiliation{%
  \institution{Technion - Israel Institute of Technology}
  \city{Haifa}
  \country{Israel}
}

\renewcommand{\shortauthors}{Madmon, Pipano, Reinman \& Tennenholtz}

\begin{abstract}
    We study a game-theoretic information retrieval model in which strategic publishers aim to maximize their chances of being ranked first by the search engine while maintaining the integrity of their original documents. We show that the commonly used Probability Ranking Principle (PRP) ranking scheme results in an unstable environment where games often fail to reach pure Nash equilibrium. We propose two families of ranking functions that do not adhere to the PRP. We provide both theoretical and empirical evidence that these methods lead to a stable search ecosystem, by providing positive results on the learning dynamics convergence. We also define the publishers' and users' welfare, demonstrate a possible publisher-user trade-off, and provide means for a search system designer to control it. Finally, we show how instability harms long-term users' welfare. 
\end{abstract}

\received{31 December 2024}
\received[revised]{15 June 2025}
\received[accepted]{24 June 2025}

\maketitle

\section{Introduction}

The rapid advancement of information retrieval systems, exemplified by search engines such as Google and Bing, and recommendation systems such as YouTube and Spotify, has reshaped how users interact with the vast digital information landscape. Central to their function is ad-hoc retrieval: ranking documents within a corpus according to their relevance to a user's query. A key phenomenon in such ecosystems is strategic behaviors from content providers, who modify their web content to improve their ranking in search results, a practice known as search engine optimization (SEO). Such strategies significantly influence the quality of search results and the user's experience. This phenomenon has garnered substantial attention, and game-theoretic tools have been applied to analyze it. Since a ranking scheme induces a game, the task of choosing a ranking scheme is essentially a task of mechanism design, a central field in game theory.

One key challenge regarding these ranking-induced games is the design of \emph{stable ranking mechanisms}, under which natural dynamics of content providers (hereafter, publishers) converge to a stable state, in which no publisher is willing to modify their content given the other publishers' content. This stable state is widely known as \emph{Nash equilibrium}, and one popular family of learning dynamics is the class of \emph{better response dynamics}, which assumes that at each round one publisher modifies her content in a way that increases her utility compared to the previous round. Note that this minimal assumption captures a wide range of behavioral patterns that are likely to be observed in practice.
A search engine designer is typically interested in a ranking scheme that guarantees the convergence of actual dynamics induced by strategic content modification in the text space. While the structure of such dynamics is unknown, a stronger approach we adopt is to design a ranking scheme that guarantees the convergence of \emph{any} better response dynamics. This approach in particular ensures the convergence of any real-life dynamics.

In this work, we study stable mechanisms for search and exposure-based recommendation systems in a continuous action space model. We adopt a framework that models the strategy space of publishers (the documents they create) as a continuous embedding space, distinguishes between publishers' and users' welfare, and accounts for publishers' initial documents, which can be thought of as preferred content that the publishers aim to maintain some level of integrity to. An important component of the model is the \emph{information need}, which is the user's underlying goal or requirement for information. We assume a dense retrieval model, in which the information need is represented in the same continuous embedding space as the publishers' documents \parencite{mitra2017neural}.

Within this model, we study the convergence of better response dynamics under different ranking schemes. Importantly, while we use the terminology of a search engine that determines publishers' exposure through an underlying ranking mechanism, our approach is agnostic to the specific ranking process. Instead, we assume that the system can directly allocate exposure to each publisher. This flexibility allows our model to extend beyond the context of information retrieval, making it applicable to exposure-based recommendation systems more broadly.

The \textbf{probability ranking principle (PRP)} is a widely used principle that states that documents should be ranked in order of decreasing probability of relevance to the user \cite{robertson1977probability}. In our model, this translates into the principle of greedily allocating the entire exposure mass to the publisher who provides the most relevant content (in terms of distance from the information need). We begin our analysis by providing a negative result on the stability of the PRP ranking scheme (Observation \ref{observation: prp example}).
We then introduce two alternative ranking schemes, the \textbf{softmax ranking function} (also discussed in \textcite{hron2022modeling}) and the \textbf{linear ranking function}.

The softmax ranking function serves as a smoothed version of the PRP ranking function, with smoothing controlled by an inverse temperature hyperparameter. We observe that the softmax ranking function is part of a broad and natural class of \emph{proportional ranking functions}. However, we also establish that none of these ranking functions induce games with an exact potential, a primary tool for proving and analyzing the convergence of better response dynamics (Theorem \ref{theorem: prop not potential}). While the absence of potential is not necessarily enough to preclude convergence of better-response dynamics, we also show an instance of the publishers' game in which softmax ranking can induce cyclic content modification patterns, thereby preventing convergence (Observation \ref{obs:softmax do not converge}).\footnote{Throughout the paper, we use the term ``potential'' to refer to an \emph{exact potential}, which is indeed only a sufficient condition for proving better response dynamics convergence. In general games, an exact potential might not exist, and better response dynamics will still be guaranteed to converge (for instance, there might exist a \emph{generalized ordinal potential}, a weaker potential notion that also guarantees convergence of learning dynamics). We show this is not the case when using the softmax ranking function.}

This contrasts with the linear ranking function, which applies a linear transformation (with the slope as a hyperparameter) to the publishers' \emph{linear relative relevance}—defined as the difference between a publisher's proximity to the information need and the average proximity of their competitors.
For this family of ranking functions, we prove the convergence of \emph{any} better response dynamics to a stable state using a potential function argument (Theorem \ref{theorem: linear potential}). We also provide a detailed equilibrium characterization and welfare-in-equilibrium analysis under the linear ranking function, utilizing its convenient theoretical properties (Theorem \ref{theorem: linear single pne + dominant}, Lemma \ref{lemma: linear PNE} and additional discussion in Appendices \ref{ape: euclidean norm} and \ref{ape: lin analysis}). Lastly, we show that it is impossible to design a valid ranking function by taking a non-linear transformation of the linear relative relevance, which justifies our design choice of linearity (Theorem \ref{theorem: linear only simplex}).

Our theoretical analysis is followed by a rigorous empirical evaluation of all three ranking functions in terms of users' and publishers' welfare.\footnote{Code and results are available in \url{https://github.com/ireinman/The-Search-for-Stability}.} We perform comparisons both between ranking families and within ranking families. Our empirical analysis demonstrates how the instability of the PRP ranking function harms its performance compared to the alternatives we suggest.
While indeed there are no theoretical guarantees for the softmax function, we empirically show it is stable--in the sense that learning dynamics converge in almost any instance--and hence manages to surpass the PRP in terms of users' welfare. This underscores the practical effectiveness of the softmax ranking function.
Additionally, we show an inherent trade-off between publishers' and users' welfare and provide a means for a system designer to control it through hyperparameter tuning.

\paragraph{Paper organization}
In \S\ref{sec: related work} we review related work in the field of strategic information retrieval. \S\ref{sec: preliminaries} provides preliminary definitions and results from game theory. In \S\ref{sec: the model} we discuss the publishers' game model. \S\ref{sec: ranking functions} provides a theoretical analysis of learning dynamics in our model, and studies stability under different ranking schemes (PRP, softmax, and linear). 
In \S\ref{sec: empirical} we provide experimental results of simulating learning dynamics under the different ranking schemes. We then conclude and present future work directions in \S\ref{sec: discussion}. The Appendix includes further theoretical developments, additional empirical results, and proof segments omitted from the main article.

\section{Related Work}
\label{sec: related work}

Our work lies in the intersection of game theory, economics, and machine learning. There is extensive literature on strategic and societal aspects of machine learning, such as classification and segmentation of strategic individuals \cite{hardt2016strategic,nissim2018segmentation,levanon2022generalized,nair2022strategic}, fairness in machine learning \cite{chierichetti2017fair,abbasi2021fair,ben2021protecting}, and recommendation systems design \cite{ben2018game,ben2019recommendation,bahar2020fiduciary,mladenov2020optimizing}. Specifically, we contribute to the study of learning dynamics within machine learning, data-driven environments, and ad auctions \cite{claus1998dynamics,freund1999adaptive,meir2010convergence,cary2014convergence}.

In this work, we study a game-theoretic model of the search (or more generally, recommendation) ecosystem, in which strategic content providers compete to maximize impressions and visibility.
\textcite{ben2015probability} and \textcite{basat2017game} showed that the widely used PRP ranking scheme of \textcite{robertson1977probability} leads to sub-optimal social welfare, and demonstrated how introducing randomization into the ranking function can improve social welfare in equilibrium. \textcite{raifer2017information} provided both theoretical and empirical analysis of a repeated game, and revealed a key strategy of publishers:  
gradually changing their documents to resemble those ranked highest in previous rounds.
\textcite{ben2019convergence} studies better-response dynamics in a publishers' game in which the strategy space is modeled as a finite set of ``topics''.
\textcite{kurland2022competitive} advocates for the devising of ranking functions that are optimized not only for short-term relevance effectiveness but also for long-term corpus effects.

While these works rely on representing documents as a finite set of ``topics'',
a more recent line of work takes a different modeling assumption of representing content as a continuous, high-dimensional space. \textcite{hron2022modeling} studies both PRP and softmax-based ranking functions, but does not incorporate initial documents, which are crucial in our analysis. \textcite{jagadeesan2023supply} studies a model similar to \textcite{hron2022modeling} but also incorporates quality constraints--the publisher’s cost is a function of the document’s norm. This is a private case of our model, in which all publishers have the same initial document. Our model generalizes by assuming publishers’ optimal documents may be different, which is a rather realistic scenario. 

In a similar spirit, \textcite{yao2024human} studies competition between human content creators and Generative AI, and \textcite{yao2024user} provides novel techniques for optimizing user welfare in equilibrium.
Importantly, none of these papers study learning dynamics or their convergence.
Closer to our work, \textcite{yao2023rethinking} studies the design of stable recommendation mechanisms under 
similar modeling analysis as in our paper. However, their work concerns the design of a reward mechanism (which can be seen as designing the ranking and payments jointly), while in our model, the designer can only determine the exposure of each publisher, without being able to transfer money. \textcite{yao2023bad} and \textcite{madmon2024convergence} also studied content provider dynamics, and adopted modeling assumptions similar to ours. 
However, their work focuses on regret minimization dynamics, where publishers act simultaneously and aim to minimize their cumulative regret, defined as the gap between the utility they could have achieved with full knowledge of others’ actions and the utility they actually obtain. In contrast, we study better-response dynamics, where at each time step, a single publisher updates its content to myopically improve its own utility.

\section{Preliminaries}
\label{sec: preliminaries}

In this section, we provide some standard definitions and results from the field of game theory. We start with the definition of a \textbf{game}:

\begin{definition}
    A game is a tuple $ \bigl(N, (X_i)_{i \in N}, (u_i)_{i \in N} \bigr)$, where $N \coloneqq \{ 1, ..., n\}$ is the set of players, and for each player $i \in N$, $X_i$ is the set of actions. The set $X \coloneqq X_1 \times ... \times X_n$ is called the set of pure strategy profiles. Each player $i \in N$ has a utility function $u_i: X \to \R$. 
\end{definition}

In the sequel, we assume that $X_i$ is a compact and convex set and that $u_i$ is a bounded function. We also assume that all players are rational, in the sense that each player attempts to maximize her own utility. Each player's utility depends not only on her action but also on the actions taken by all other players. We denote by $X_{-i} \coloneqq X_1 \times ... \times X_{i-1} \times X_{i+1} \times ... \times X_n$ the set of all pure strategy profiles of all players but player $i$, and use the notation of $x_{-i} \in X_{-i}$ to denote such a strategy profile of everyone but $i$. An important concept is the concept of \textbf{$\varepsilon$-better response}.

\begin{definition}
    Given a strategy profile $x_{-i} \in X_{-i}$ of all players but $i$ and actions $x_i, \; x'_i \in X_i$ of player $i$, we call $x'_i$ an $\varepsilon$-better response than $x_i$ with respect to $x_{-i}$ if $u_i(x'_i, x_{-i}) > u_i(x_i, x_{-i}) + \varepsilon$.
\end{definition}

The notion of $\varepsilon$-best response allows defining the solution concept of an \textbf{$\varepsilon$-pure Nash equilibrium}: a pure strategy profile from which no player can improve her utility by more than $\varepsilon$ by deviating (i.e., changing her actions given all other players' actions are fixed).

\begin{definition}
    A pure strategy profile $x \in X$ is called an $\varepsilon$-pure Nash equilibrium ($\varepsilon$-PNE), if for every player $i \in N$, \textbf{no} $x'_i$ is an $\varepsilon$-better response than $x_i$ with respect to $x_{-i}$.
\end{definition}

For $\varepsilon=0$, $0$-better response and $0$-PNE are called better response and PNE, respectively. A PNE is often considered a stable state of a game \cite{ben2018game}.
When players apply the concept of $\varepsilon$-better response we get the following type of dynamics:

\begin{definition}
    An \textbf{$\varepsilon$-better response dynamics} is a sequence of profiles $\bigl \{ x^{(t)} \bigr \}_{t=1}^{\infty}  \subseteq X$ such that in each timestep $t$, either $x^{(t-1)}$ is an $\varepsilon$-PNE and $x^{(t)}=x^{(t-1)}$, or there is a single deviator $i \in N$ such that $x^{(t)}_{-i} = x^{(t-1)}_{-i} $ and  $x^{(t)}_{i}$ is an $\varepsilon$-better response than $x^{(t-1)}_{i}$ with respect to $x^{(t-1)}_{-i}$. 
    We say such a dynamics has \textbf{converged} if there exists a timestep $T$ such that $x^{(T)}$ is an $\varepsilon$-PNE.
\end{definition}

Another class of games, which will be of interest in this paper, is the class of \textbf{potential games}.

\begin{definition}[\citeauthor{monderer1996potential}]
    A game $G$ is called a potential game if there exists a \emph{potential} $\phi: X \to \R$, such that for any $i \in N, x_{-i} \in X_{-i}, x'_i, x''_i \in X_i$: \\
    \begin{equation}
        \phi(x'_i, x_{-i}) - \phi(x''_i, x_{-i}) = u_i(x'_i, x_{-i}) - u_i(x''_i, x_{-i})
    \end{equation}
\end{definition}

That is, in a potential game, a single function $\phi$ ``captures'' the incentives of all the players. It is straightforward that any potential maximizer is a PNE. Moreover, \textcite{monderer1996potential} shows a significant result:

\begin{theorem}[\citeauthor{monderer1996potential}]
\label{theorem: BR converges}
    Let $G$ be a potential game and let $\varepsilon > 0$. Then, $G$ possesses at least one $\varepsilon$-PNE, and any $\varepsilon$-better response dynamics in $G$ converges.
\end{theorem}

Another important concept is the concept of a \textbf{strictly dominant strategy}, a strategy which is the best choice for a player regardless of the strategies chosen by other players. The significance of strictly dominant strategies lies in their stability and predictability. 

\begin{definition}
    A strategy $x_i\in X_i$ is a strictly dominant strategy for player $i\in N$ if $\forall x_{-i}\in X_{-i}, x'_i\in X_i$, $u_i(x_i, x_{-i}) > u_i(x'_i, x_{-i})$.
\end{definition}

\section{The Model}
\label{sec: the model}

We start by defining the model of a publishers' game. A publishers' game consists of $n$ publishers (players), each can provide content in the continuous $k$ dimensional embedding space $[0,1]^k$, where the distance between documents is measured by some continuous semi-metric $d$.\footnote{A semi-metric is a function that satisfies all the axioms of a metric, with the possible exception of the triangle inequality.} This could be, for example, the Euclidean distance, which is widely used in the context of embedding \cite{schroff2015facenet,agrawal2021minimum,alshamrani2024distance}.

Each publisher $i$ has a preferred content type (or: initial document), denoted by $x_0^i \in [0,1]^k$. We denote by $x^* \in [0,1]^k$ the information need representation in the feature space. The term $x^*$ can refer to a particular query or topic in the search environment.\footnote{Alternatively, when considering a recommender system that places items (documents) and users in the same latent space, $x^*$ can be thought of as a representative user.}
In our model, the information need is represented in the same embedding space as the documents, inspired by BERT rankers and dense retrieval models \cite{qiao2019understanding,zhan2020learning,zhao2024dense}.
While the exact information need is typically latent, it is standard in the information retrieval literature to assume that it can be modeled as a point in a semantic or topical space \cite{chowdhury2010introduction}.

We assume the information need \(x^*\) is known to the publishers. While this assumption may seem restrictive, we claim that it is justified, for example, in the setting of a \emph{niche domain}, where typically a small number of publishers compete over a specific domain. In such scenarios, the publishers specialize in publishing content within a specific domain and know precisely the users’ preferences and needs.
Examples for such domains include \emph{travel content}, where publishers generate destination-specific articles tailored to well-understood traveler interests, and \emph{professional content creators} who produce material for clearly defined audiences in domains in which the publishers are often experts themselves (such as finance or sports).

The search engine uses a ranking function $r$ to determine the documents' ranking. We assume that $r$ is oblivious, meaning it is independent of the publishers' identities. The choice of $r$ induces a strategic interaction among the publishers, which can be formally defined as a game:

\begin{definition}
    A \textbf{publishers' game} $G$ is a tuple \\
    $(N, k, \lambda, d, (x_0^i)_{i \in N}, x^*, r)$, where:
    \begin{itemize}
        \item $N \coloneqq \{ 1, 2, ... n \}$ is the set of publishers, with $n \ge 2$;
        \item $k$ is the embedding-space dimension of documents;
        \item $X_i \coloneqq [0,1]^k$ is the set of publisher $i$'s possible actions;
        \item $\lambda > 0$ is the factor of the cost for providing content that is different from the initial document;
        \item $d: [0,1]^k \times [0,1]^k \to \R_{+}$\footnote{We use $\R_{+}$ to denote the set of non-negative real numbers and $\R_{++}$ to denote the set of strictly positive real numbers.} is a continuous semi-metric satisfying $\forall x, y: d(x,y) \le 1$;\footnote{This is without loss of generality, as for any distance function that does not satisfy $\max_{x,y} d(x,y) \le 1$, one can simply divide it by the maximal distance.}
        \item $x_0^i \in X_i$ is publisher $i$'s initial document;
        \item $x^*\in [0,1]^k$ is the information need;
        \item $r: X \to \Delta^n$ is a ranking function, that might depend on the information need $x^*$.
    \end{itemize}
\end{definition}

The ranking function $r$ determines the distribution over the winning publishers given a strategy profile $x \in X$, with the interpretation that $r_i(x)$ is the probability of publisher $i$ being ranked first by $r$ under the strategy profile $x$. This modeling assumption is motivated by the empirical evidence that users usually prefer to reformulate their query rather than searching beyond the first retrieved result \cite{butman2013query,liu2016impacts,joachims2017accurately}. We abuse the notation and denote $d^*(x_i) \coloneqq d(x_i, x^*)$ and $d^0_i(x_i) \coloneqq d(x_i, x_0^i)$.
The utility of publisher $i$ under the profile $x$ is defined to be:

\begin{equation}
    u_i(x) \coloneqq r_i(x) - \lambda d^0_i(x_i)
\end{equation}

That is, the utility of publisher $i$ is the probability of winning minus the cost of providing content that differs from her initial document. 
We now define the \textbf{publishers' welfare} $\mathcal{U}(x)$ and the \textbf{users' welfare} $\mathcal{V}(x)$, which will henceforth be referred to collectively as \textbf{welfare measures}.
The publishers' welfare is the sum of the publishers' utilities, while the users' welfare is the expected relevance of the first-ranked document, where we quantify the relevance of a document $x_i$ by $1-d^*(x_i)$.  
\begin{gather}
    \mathcal{U}(x) \coloneqq \sumi u_i(x) = 1 - \lambda \sumi d^0_i (x_i) \\
    \mathcal{V} (x) \coloneqq \sumi \bigl (1 - d^*(x_i) \bigr ) \cdot r_i(x) = 1 - \sumi d^*(x_i) \cdot r_i(x)
\end{gather}

\begin{example}[Running example]\label{example:model}
    To illustrate our model, we consider a simple example involving two publishers competing for exposure in response to a specific information need. In this toy example, the embedding space is two-dimensional ($k = 2$), and the distance metric is the squared Euclidean distance, defined as $d(x, y) = \frac{1}{2} \|x - y\|_2^2$.
    
    The publishers compete over a population of users whose information need can be described as \emph{adventurous eco-tourism}. Let $x^*=(0.358,0.908)$ be the embedding representation of the information need. While many small independent publishers may produce content relevant to this topic, we assume that, within this niche domain, there are only two prominent publishers, receiving the vast majority of exposure, with all other publishers receiving negligible attention in comparison. We therefore only consider the two prominent publishers in this running example.
    
    We assume that both publishers share a cost parameter of $\lambda = 1$.
    Publisher 1 specializes in \emph{Arctic wildlife} and her initial document embedding vector is given by $x_1^0 = (0.950,0.035)$, while
    Publisher 2 focuses on \emph{tropical hiking trails}, having $x_2^0 = (0.933,0.773)$ as her initial document representation. 
\end{example}

We will refer to this running example throughout Section~\ref{sec: ranking functions} to illustrate the behaviors induced by different ranking functions, highlighting properties such as the existence and characterization of PNE, as well as the convergence (or lack thereof) of learning dynamics.

\section{Ranking Functions and Learning Dynamics}
\label{sec: ranking functions}

This section presents a theoretical analysis of the learning dynamics in our model and studies stability under three ranking functions: the PRP ranking function, the softmax ranking function, and the linear ranking function.

\subsection{The PRP Ranking Function}

The Probability Ranking Principle (PRP; \cite{robertson1977probability}) is a key principle in information retrieval, which forms the theoretical foundations for probabilistic information retrieval. The principle says that documents should be ranked in decreasing order of the probability of relevance to the information need expressed by the query. In our model, the analogous principle means that the closest document to the information need should always be ranked first by the ranking function. We begin by formally defining the PRP ranking function in our model:

\begin{definition}
    The \textbf{PRP ranking function}, denoted by $r^*$, is the ranking function defined as follows:
    \begin{equation}
        r_i^*(x) \coloneqq \frac{1}{|\mu(x)|}  \cdot \mathds{1}_{i \in \mu(x)} 
    \end{equation}
    where $\mu(x) \coloneqq \argmin_{i \in N} d^*(x_i)$ is the set of all publishers providing the closest content to the information need $x^*$.
\end{definition}

Note that the PRP yields optimal short-term retrieval in the sense that it maximizes the users' welfare $\mathcal{V}(x)$ for a fixed strategy profile $x$. However, its discontinuity induces unstable ecosystems, in which there might not exist a PNE. This instability causes sub-optimal long-term effects, as we show later. 

\begin{example}\label{example:prp}

    Consider the instance presented in Example \ref{example:model}, with the PRP as its ranking function. Now, consider the strategy profile in which $x_1 = x_2 = x^*$, namely, both publishers precisely target the most suitable content given the information need. Under this profile, the utility of Publisher 1 is given by:
    
    \begin{gather}
        \begin{aligned}
            u_1(x_1, x_2) & = r_1(x_1, x_2) - \lambda d^0_1(x_1) \\
            &= r_1(x^*, x^*) - 1 \cdot \frac{1}{2}\norm{(0.950, 0.035) - (0.358, 0.908)}^2 \\
            & \approx \frac{1}{2} - 0.556 = -0.056 < 0
        \end{aligned}
    \end{gather}
    
    Now, observe that if Publisher~1 deviates and chooses $x_1 = x^0_1$, her utility becomes $0$: on the one hand, she will never be ranked first, but on the other hand, she incurs no cost for deviating from her initial document.
    Therefore, the profile of $x_1 = x_2 = x^*$ is not a PNE, as Publisher 1 has a beneficial deviation.

\end{example}

In Example~\ref{example:prp}, we showed that the profile in which both players choose $x^*$ is not a PNE. In fact, the following observation demonstrates that in many publishers' games induced by the PRP ranking function, no pure Nash equilibrium exists at all.

\begin{observation}
\label{observation: prp example}
Let $k\in \N$ and let $G$ be a publishers' game with $n=2$, $\lambda > \frac{1}{2}$, $d(x,y)=\frac{1}{k}\norm{x-y}^2_2$, $r=r^*$, $x_0^1=x_0^2=\vec{0}_k$ and $x^* = \vec{1}_k$. Then, $G$ possesses no PNE.
\end{observation}

\begin{proof}
    Assume by contradiction that $x^{eq}\in X$ is a PNE of $G$. We make use of the following auxiliary lemma (whose proof is given in Appendix \ref{ape: proofs}):

    \begin{lemma}\label{lemma: prp trivial pne}
        Let $G$ be a publishers' game with $r = r^*$ and $d(x,y)=\frac{1}{k}\norm{x-y}_2^2$.
        If $x^{eq}$ is a PNE of $G$, then $x^{eq}_i \in \{ x^*, x_0^i\} \; \forall i \in N$.
    \end{lemma}

    By Lemma \ref{lemma: prp trivial pne}, there are only 4 profiles that can be PNEs of $G$. We proceed to show that none of them is a PNE.
    \begin{itemize}
        \item $x^{eq} = (x^*, x_0^2)$: player 1 can improve her utility by deviating to
         $\hat{x}_1 = (\frac{1}{2},..., \frac{1}{2})$.
        \item $x^{eq} = (x_0^1, x^*)$: player 2 can improve her utility by deviating to
         $\hat{x}_2 = (\frac{1}{2},..., \frac{1}{2})$.
        \item $x^{eq} = (x_0^1, x_0^2)$: player 1 can improve her utility by deviating to $\hat{x}_1 = (\alpha, ..., \alpha)$ for $0 < \alpha < \sqrt{\frac{1}{2\lambda}} $ since $u_1(\hat{x}_1, x^{eq}_{-1}) = 1 - \lambda \alpha^2 > 1 - \lambda \frac{1}{2\lambda} = \frac{1}{2} = u_1(x^{eq})$.
        \item $x^{eq} = (x^*, x^*)$: player 1 can improve her utility by deviating to $x_0^1$ since $u_1(x_0^1, x^{eq}_{-1}) = 0 - 0 > \frac{1}{2} - \lambda = u_1(x^{eq})$. 
    \end{itemize}
\end{proof}

As we discuss throughout the paper, the presence of a PNE holds significant importance for search engine designers.
We now turn to suggest two other ranking function families, which will be shown (either theoretically or empirically) to be stable in terms of PNE existence and $\varepsilon$-better response dynamics convergence.

\subsection{The Softmax Ranking Functions}

Note that the instability of the PRP arises from the discontinuity of the argmax function, which introduces discontinuities into the ranking itself.\footnote{It is actually the \(\argmin\) that is used by the PRP, but $\argmin_{i \in N} d^*(x_i) = \argmax_{i \in N} \{- d^*(x_i)\}$.}
Hence, a natural way to construct a smooth version of the PRP would be to replace argmax with its well-known smooth variant: the softmax function. Using softmax as a replacement for argmax is a standard practice in machine learning, where it serves as a differentiable approximation to discrete choices. This approach was successfully applied to strategic classification \cite{levanon2022generalized}, item representation learning \cite{nahum2023decongestion}, information design \cite{yu2023encoding}, and beyond. In our context, the softmax ranking function is naturally defined as follows.\footnote{Interestingly, the publishers' game induced by the softmax ranking function generalizes the well-known rent-seeking competition of \textcite{hirshleifer1989conflict}.}

\begin{definition}
    The \textbf{softmax} ranking function with inverse temperature constant $\beta > 0$, denoted by $\tilde{r}^{\beta}$, is the ranking function defined as follows:
    \begin{equation}
            \tilde{r}_i^{\beta}(x) \coloneqq \frac{e^{-\beta \cdot d^*(x_i)}}{\sumj e^{-\beta \cdot d^*(x_j)}}
    \end{equation}
\end{definition}

Note that as $\beta \to \infty$, $\tilde{r}^{\beta} \to r^*$. This further explains the intuition behind the usage of the softmax function in our context and shows that $\beta$ is a means to control how closely $\tilde{r}^{\beta}$ aligns with the PRP.
The class of softmax ranking functions is a subset of a wider class of ranking functions, known as the class of proportional ranking functions \cite{madmon2024convergence}:

\begin{definition}
    We say a ranking function $r$ is \textbf{proportional} if there exists a continuously differentiable function $g:[0,1] \to \R_{++}$ such that
    \begin{equation}
        r_i(x) = \frac{g \bigl (d^*(x_i) \bigr )}{
        \sumj g \bigl (d^*(x_j) \bigr )}
    \end{equation}
\end{definition}

The proportionality condition is an intuitive way to construct a function whose image is contained in the simplex $\Delta ^n$. We say a ranking function $r$ \textbf{induces} a potential game if any publishers' game with $r$ is a potential game. The following theorem shows that, despite its intuitive appeal, the class of proportional ranking functions (including the softmax ranking functions) is fundamentally incompatible with the desirable structure of potential games.

\begin{theorem}
\label{theorem: prop not potential}
    Except for the uniform ranking function \(r_i \equiv \frac{1}{n} \), there exists no proportional ranking function that induces a potential game.
\end{theorem}

\begin{proof}
    Assume by contradiction that there exists a proportional ranking function $r$, defined by $r_i(x) = \frac{g \bigl (d^*(x_i) \bigr )}{\sumj g \bigl (d^*(x_j) \bigr )}$ 
    for some continuously differentiable \(g:[0,1] \to \R_{++}\), which is both not the uniform ranking function (\(r_i \equiv \frac{1}{n}\)) and induces a potential game. We show that the fact that $r$ induces a potential game implies that $g$ is constant, which implies \(r\) is the uniform ranking function and completes the proof.

    Since $r$ induces a potential game, any publishers' game with $r$ as its ranking function is a potential game.
    In particular, any publishers' game with $r$ as its ranking function,
    $k=1$ and $d$ twice continuously differentiable is a potential game.
    Such games, since $g > 0$, satisfy the conditions
    of Theorem 4.5 in \textcite{monderer1996potential}.
    Therefore:
    \begin{equation}
        \frac{\partial^2 u_i (x)}{\partial x_i \partial x_j} =
        \frac{\partial^2 u_j (x)}{\partial x_i \partial x_j}
        \;\;\; \forall i,j\in [n], \forall x\in X
    \end{equation}
    Recall $u_i(x) \coloneqq r_i(x) - \lambda d_i^0(x_i)$. Since $\frac{\partial^2 \lambda d_i^0 (x_i)}{\partial x_i \partial x_j} = \frac{\partial^2 \lambda d_j^0 (x_i)}{\partial x_i \partial x_j} = 0$, we get:
    \begin{equation}
        \frac{\partial^2 r_i (x)}{\partial x_i \partial x_j} =
        \frac{\partial^2 r_j (x)}{\partial x_i \partial x_j}
        \;\;\; \forall i,j\in [n], \forall x\in X
    \end{equation}
    Note that $r_i$ can be written as:
    \begin{equation}
        r_i(x) = \frac{(g\circ d^*)(x_i)}{\sum\limits_{j=1}^n(g\circ d^*)(x_j)}
    \end{equation}
    
    We now make use of the following lemma (whose proof is given in Appendix \ref{ape: proofs}):
    \begin{lemma}
\label{lemma: second order condition}
    Let $f:\mathbb{R}^n \to \mathbb{R}^n$ 
    be of the form $f_i(x) = \frac{g(x_i)}{\sum\limits_{j=1}^n g(x_j)}$ for a continuously differentiable function $g:[0,1]\to\mathbb{R}$.
    Then 
    \begin{equation}\label{double derivatives}\frac{\partial^2 f_i (x)}{\partial x_i \partial x_j} =
    \frac{\partial^2 f_j (x)}{\partial x_i \partial x_j}
    \;\;\; \forall i,j \in [n], \forall x\in [0,1]^n
    \end{equation}
    iff $g$ is constant on $[0,1]$.
\end{lemma}

    By Lemma \ref{lemma: second order condition}, $g\circ d^*$ is constant on $[0,1]$.
    And since this is true for every twice continuously
    differentiable $d^*$, it is true in particular for $x^*=0, d(a,b)=(a-b)^2$, which gives us $d^*(a) = a^2$. Note that in this case, the image of $d^*$ is precisely $[0,1]$.
    Therefore, $g$ is constant on $[0,1]$.

\end{proof}   

Theorem \ref{theorem: prop not potential} implies, in particular, that the softmax ranking function does not induce a potential game.
One might naturally question whether the convergence of better response dynamics under the softmax ranking function is still guaranteed, despite the game not being a potential game (for example, there might exist a generalized ordinal potential). However, Observation \ref{obs:softmax do not converge}, which uses the running example, demonstrates that this is not the case. While Observation \ref{obs:softmax do not converge} demonstrates this for a specific value of $\beta$, we hypothesize that for any given $\beta$, one can construct a game instance for which better-response dynamics diverge under the softmax ranking function with inverse temperature $\beta$.

\begin{observation}
\label{obs:softmax do not converge}
There exist $\beta > 0$, a publishers' game instance $G$ with $r = \tilde{r}^{\beta}$ and $\varepsilon > 0$ such that $\varepsilon$-better response dynamics may not converge.
\end{observation}
\begin{proof}
Consider the instance from Example \ref{example:model}, with the softmax ranking function with $\beta = 5$.
Now, consider the following four strategy profiles ($i$'th row is publisher $i$'s strategy):
    \begin{equation}
        x^{(1)} = \begin{pmatrix}
            0.62959224 & 0.5189262 \\
            0.61146602 & 0.8503398
        \end{pmatrix}, \quad x^{(2)} = \begin{pmatrix}
            0.62959224 & 0.5033012 \\
            0.61146602 & 0.8503398
        \end{pmatrix}
    \end{equation}

    \begin{equation}
        x^{(3)} = \begin{pmatrix}
            0.62959224 & 0.5033012 \\
            0.62251457 & 0.83929126
        \end{pmatrix}, \quad x^{(4)} = \begin{pmatrix}
            0.62959224 & 0.5189262 \\
            0.62251457 & 0.83929126
        \end{pmatrix}
    \end{equation}

Notice that $x^{(2)}$ is a unilateral deviation of Publisher $1$ from $x^{(1)}$, $x^{(3)}$ is a unilateral deviation of Publisher $2$ from $x^{(2)}$, $x^{(4)}$ is a unilateral deviation of Publisher $1$ from $x^{(3)}$, and $x^{(1)}$ is a unilateral deviation of Publisher $2$ from $x^{(4)}$. It is straightforward to show that each such deviation strictly improves the utility of the deviating publisher, and therefore, for a sufficiently small $\varepsilon > 0$, the four strategy profiles define a cyclic (hence not converging) $\varepsilon$-better response dynamics.

\end{proof}
 
Such instability naturally becomes more likely as $\beta$ increases, causing the game to resemble the one induced by the PRP ranking function. However, as demonstrated in our empirical evaluation, there exists a range of $\beta$ values where the softmax ranking function exhibits a remarkable degree of stability across nearly all game instances. Furthermore, for certain $\beta$ values, the user welfare in the reached equilibrium surpasses the average user welfare achieved under the PRP ranking. This implies that, from a purely practical perspective, a designer can fine-tune the selection of $\beta$ to achieve both user welfare and stability.
Nevertheless, the lack of a theoretical guarantee regarding learning dynamics convergence motivates the study of linear ranking functions, presented hereinafter.

\subsection{The Linear Ranking Functions}
\label{linear_theory}

We begin by defining the linear relative relevance, which is the difference between the average distance of all other publishers from $x^*$ and the distance of publisher $i$ from $x^*$.

\begin{definition}
    Let $x \in X$ be a strategy profile in a publishers' game. The \textbf{linear relative relevance} of document (or publisher) $i \in N$ with respect to $x$ is:
    \begin{equation}
            \nu_i(x) \coloneqq \frac{1}{n-1} \Bigl( \sum_{j \in N \setminus \{i\}} d^*(x_j) \Bigr) - d^*(x_i)
    \end{equation}
\end{definition}

Note that player \(i\)'s linear relative relevance increases as she becomes more relevant to the information need (minimizes \(d^*(x_i)\)) and decreases as any other publisher does so.
A natural way to construct a ranking function using \(\nu\) would be expressions like \(\hat{r}^{a,b}_i(x) \coloneqq a \cdot \nu_i(x) + b\).
Importantly, \(\sum_{i=1}^{n} \nu_i \equiv 0\).
Hence \(b\) must equal \(\frac{1}{n}\) for the image of \(\hat{r}^{a,b}\) to be contained in \(\Delta^n\).
Further note that \(\nu_i(x)\) might get the value \(-1\) so we must have \(a \leq \frac{1}{n}\) to ensure \(a \cdot \nu_i(x) + \frac{1}{n} \ge 0\).
Lastly, since a publisher with high \(\nu_i\) should get more exposure, \(a\le0\) would not be plausible. This gives rise to the following definition:

\begin{definition}
    The \textbf{linear} ranking function with slope \(a \in (0, \frac{1}{n}]\), denoted by $\hat{r}^{a}$, is the ranking function defined as follows:
    \begin{equation}
        \hat{r}^{a}_i (x) \coloneqq a \cdot \nu_i(x) + \frac{1}{n}
    \end{equation}
\end{definition}

A shortcoming of the linear ranking functions is evident from the above definition – the parameter $a$ needs to be quite small as $n$ goes large. This is problematic because it renders the value of $r$ in a very small range, restricting the extent to which the ranking function incentivizes publishers to produce relevant content. However, as we highlighted when presenting the model, the case of small $n$ is well-motivated, for example, in the case of niche domains. In such scenarios, the condition \(a\in(0,\frac{1}{n}]\) is not particularly restrictive.\footnote{Additionally, in a real-life scenario, such a ranking function can incorporate an initial filtering of the documents based on a certain distance threshold, and only afterward apply the linear ranking function to determine the probability of the remaining documents being ranked first.}

We now proceed to provide some results about the significant advantages of the linear ranking function. Importantly, all linear ranking functions induce a potential game:

\begin{theorem}
\label{theorem: linear potential}
    Let $a \in (0, \frac{1}{n}]$. $\hat{r}^a$ induces a potential game, and the following is a potential function: 
    \begin{equation}
        \hat{\phi}(x) = - \sum_{i=1}^n \bigr( a \cdot d^*(x_i) + \frac{1}{n} + \lambda d^0_i(x_i) \bigl)
    \end{equation}
\end{theorem}

\begin{proof}
Let $g(t) = a t + \frac{1}{n}$, and notice that the linear ranking function can now be written as $\hat{r}_i(x) = g(\nu_i(x))$. We will show that the following function is a potential function for the game $G$:
\begin{equation}
    \phi(x) = - \sum_{i=1}^n \bigr( -g(d^*(x_i)) + \lambda d^0_i(x_i) \bigl)
\end{equation}
To show this, one should show that for any $i \in N, x_{-i} \in X_{-i}, x'_i, x''_i \in X_i$:
\begin{equation}
\label{required}
    \phi(x'_i, x_{-i}) - \phi(x''_i, x_{-i}) = u_i(x'_i, x_{-i}) - u_i(x''_i, x_{-i})
\end{equation}

Let $i \in N, x_{-i} \in X_{-i}, x'_i, x''_i \in X_i$. Starting with the LHS, we get:
\begin{equation}
\label{lhs}
    \begin{aligned}
        \phi(x'_i, x_{-i}) - \phi(x''_i, x_{-i}) & = 
        - g(d^*(x'_i)) - \lambda d^0_i(x'_i) + g(d^*(x''_i)) + \lambda d^0_i(x''_i) \\
        & = g(d^*(x''_i) - d^*(x'_i)) + \lambda(d^0_i(x''_i) - d^0_i(x'_i)) - \frac{1}{n}
    \end{aligned}
\end{equation}

On the other hand, simplifying the RHS yields:
\begin{equation}
\label{rhs}
    \begin{aligned}
        u_i(x'_i, x_{-i}) - u_i(x''_i, x_{-i}) & = 
        g(\nu_i(x'_i, x_{-i})) - \lambda d^0_i(x'_i) - g(\nu_i(x''_i, x_{-i})) + \lambda d^0_i(x''_i) \\
        & = g(\nu_i(x'_i, x_{-i}) - \nu_i(x''_i, x_{-i})) + \lambda(d^0_i(x''_i) - d^0_i(x'_i)) - \frac{1}{n} \\
        & = g(d^*(x''_i) - d^*(x'_i)) + \lambda(d^0_i(x''_i) - d^0_i(x'_i)) - \frac{1}{n}
    \end{aligned}
\end{equation}

The required equation \eqref{required} now follows directly from \eqref{lhs} and \eqref{rhs}, which concludes the proof.

\end{proof}

The following result follows from Theorem \ref{theorem: BR converges} and from the continuity of $\hat{\phi}$:

\begin{corollary}
\label{corollary: linear con}
    Let $a \in (0, \frac{1}{n}]$. Any publishers' game with $r = \hat{r}^a$ has at least one PNE. Moreover, for any $\varepsilon > 0$, any $\varepsilon$-better response dynamics converges.
\end{corollary}

This result has major implications for the stability of the ecosystem. Specifically, if \emph{any} $\varepsilon$-better response dynamics converges, then, in particular, \emph{any} real-life better response dynamics converges.
The result regarding learning dynamics convergence does not require full information of the initial documents of all players, but only assumes each player knows their initial document and the information need.

Note that whenever $d$ is strictly biconvex, that is, $d(x,y)$ is strictly convex in $x$ for any fixed $y$ and vice versa (which is the case for semi-metrics like squared Euclidean norm), $\hat{\phi}$ is strictly concave. Together with the unique structure of the linear ranking functions, this allows the derivation of the following:

\begin{theorem}
\label{theorem: linear single pne + dominant}
     Let $a \in (0, \frac{1}{n}]$. If $d$ is strictly biconvex, then any publishers' game with $r = \hat{r}^a$ has a unique PNE and the equilibrium strategies are strictly dominant strategies.
\end{theorem}

\begin{proof}
    
Fix some publisher $i$ and let us show she has a strictly dominant strategy. Player $i$'s utility is given by

\begin{equation}
\begin{aligned}
    u_i(x) & = r_i(x) - \lambda d_i^0(x_i) = 
    \frac{\nu_i(x)}{n} + \frac{1}{n} - \lambda d_i^0(x_i) 
    \\ & = \Bigl ( \sum\limits_{\substack{j \in N \setminus \{i\}}} \frac{d^*(x_j)}{n(n-1)} \Bigr ) - \frac{d^*(x_i)}{n} + \frac{1}{n} - \lambda d_i^0(x_i)
\end{aligned}
\end{equation}

Now, $\forall x_{-i} \in X_{-i}$:
\begin{equation}
    \begin{aligned}
    \argmax_{x_i \in X_i}\{u_i(x_i, x_{-i})\} &
     = \argmax_{x_i \in X_i}\Bigl\{ - \frac{d^*(x_i)}{n}  - \lambda d^0_i(x_i) \Bigr\}
     \end{aligned}
\end{equation}

meaning that $\argmax_{x_i \in X_i}\{u_i(x_i, x_{-i})\}$ does not depend on $x_{-i}$. Note that since $g(x) = d(x,y)$ is strictly convex for any $y \in {[0, 1]}^k$, we get in particular that $d^0_i$ and $d^*$ are strictly convex. This implies that $- \frac{d^*(x_i)}{n}  - \lambda d^0_i(x_i)$ is strictly concave in $x_i$, and therefore has a unique maximizer. So we got that $\argmax_{x_i \in X_i}\{u_i(x_i, x_{-i})\}$ is a singleton that does not depend on $x_{-i}$.
Let us denote $\argmax_{x_i \in X_i}\{u_i(x_i, x_{-i})\} = \{x^{dom}_i\}$. Then, the strategy $x^{dom}_i$ is a strictly dominant strategy for player $i$ by definition.

We proved that each player has a strictly dominant strategy. Note that the strategy profile where each player plays her strictly dominant strategy is a PNE. The uniqueness of the PNE follows from the fact that a strictly dominated strategy can never be an equilibrium strategy, and each player has only one strategy that is not strictly dominated (her dominant strategy).
\end{proof}

The theorem shows that whenever $d$ is strictly biconvex, not only do the linear ranking functions possess the strong behavioral property that any $\varepsilon$-better response learning dynamics converges, but also the resulting strategies are strictly dominant strategies for the publishers.

From now on, we focus our theoretical analysis on $\hat{r} \coloneqq \hat{r}^{\frac{1}{n}}$. The choice $a=\frac{1}{n}$ is natural, as increasing the slope incentivizes the creation of content relevant to users, but we can use the same analysis method for other slopes. The following Lemma characterizes the unique equilibrium point in the case of the squared Euclidean semi-metric, in which the equilibrium strategy of each publisher is a weighted average of her initial document and $x^*$.\footnote{In Appendix \ref{ape: euclidean norm} we provide equilibrium analysis with \(L_2\) norm.}

\begin{lemma}
\label{lemma: linear PNE}
    Any publishers' game with $r = \hat{r}$ and $d(x,y)=\frac{1}{k}||x-y||^2_2$  has a unique PNE, $x^{eq}$, given by $x^{eq}_i = \frac{x^* + \lan x_0^i}{1 + \lambda n}$. Moreover, $x^{eq}_i$ is a strictly dominant strategy for any publisher $i \in N$.\footnote{Our model can be extended to an incomplete information setting where each publisher \(i\) is aware only of \(x^*\) and of \(x_0^i\). One can show that $x^{eq}$ is an equilibrium also in such games (i.e., an ex-post equilibrium).}
\end{lemma}

\begin{proof}
Let $G$ be a publishers' game with $r = \hat{r}$ and $d(x,y)=\frac{1}{k}||x-y||^2_2$.
From Theorem \ref{theorem: linear potential}, a potential function for G is: 
\begin{equation}
    \hat{\phi}(x) = - \sum_{i=1}^n \biggr(\frac{d^*(x_i)}{n} + \frac{1}{n} + \lambda d^0_i(x_i) \biggl) = 
- \sum_{i=1}^n \biggr(\frac{||x_i-x^*||^2_2}{nk} + \frac{1}{n} + \lambda \frac{||x_i-x_0^i||^2_2}{k} \biggl)
\end{equation}

We notice that $G$ is smooth and $\hat{\phi}$ is concave, therefore, a strategy profile is a PNE if and only if it is a potential maximizer (see, for example, \cite{neyman1997correlated}).
In order to find a potential maximizer, we will find stationary points by setting the gradient of the potential with respect to $x_i$ to $0_k$. We will denote those points by $x^{eq}$.

\begin{equation}
    \nabla_{x_i} \hat{\phi}(x) = \frac{-2}{nk}(x_i-x^*) - 
    \frac{2 \lambda}{k}(x_i-x_0^i)
\end{equation}

\begin{equation}
\begin{aligned}
    0_k & = \nabla_{x_i}\hat{\phi}(x^{eq}) \implies 0_k = -(x^{eq}_i-x^*) - \lan (x^{eq}_i-x_0^i) 
    \\ & \implies 0_k = -\lan x^{eq}_i - x^{eq}_i + x^* + \lan x_0^i
   \\ & \implies \forall i \in N, \; x^{eq}_i = \frac{x^* + \lan x_0^i}{1 + \lan} 
   \end{aligned}
\end{equation}

Since $\hat{\phi}$ is strictly concave, $x^{eq}$ is its unique maximizer and hence the only PNE in $G$.
Note that the squared Euclidean norm is strictly biconvex.\footnote{That is, $g(x) = \frac{1}{k}\norm{x-y}^2_2$ is strictly convex for any $y \in {[0, 1]}^k$.} Therefore by Theorem \ref{theorem: linear single pne + dominant}, $x^{eq}_i$ is a strictly dominant strategy for every player $i$.
\end{proof}

\begin{example}\label{example:linear}
    Returning to our Example \ref{example:model}, Lemma \ref{lemma: linear PNE} can be used to find the unique PNE in the game induced by the linear ranking function. In this example, we use the maximal slope $a=\frac{1}{n}=\frac{1}{2}$.
    The unique PNE of the game, in which the strategy of each player is also a dominant strategy, is given by:
    
    \begin{gather}
        x^{eq}_1 = \frac{x^* + \lan x_0^1}{1 + \lan} = \frac{(0.358, 0.908) + 1 \cdot 2 \cdot (0.950, 0.035)}{1 + 1 \cdot 2} \approx (0.753, 0.326)
        \\
        x^{eq}_2 = \frac{x^* + \lan x_0^2}{1 + \lan} = \frac{(0.358, 0.908) + 1 \cdot 2 \cdot (0.933, 0.773)}{1 + 1 \cdot 2} \approx (0.741, 0.818)
    \end{gather}
    
    Notice that by Corollary \ref{corollary: linear con} (and by the uniqueness of the equilibrium), any better-response dynamics will converge to this profile.
    Under this equilibrium, the linear relative relevance of Publisher 1 is given by:
    
    \begin{gather}
        \begin{aligned}
            \nu_1(x^{eq}) & = \frac{1}{2-1} d^*(x^{eq}_2) - d^*(x^{eq}_1) \\& \approx \frac{1}{2}\norm{(0.741, 0.818) - (0.358, 0.908)}^2 - \frac{1}{2}\norm{(0.753, 0.326) - (0.358, 0.908)}^2
            \\& \approx 0.077 - 0.247 = -0.17
        \end{aligned}
    \end{gather}
    
    Correspondingly, the linear relative relevance of Publisher 2 is $\nu_2(x^{eq}) = - \nu_1(x^{eq}) \approx 0.017$. The utilities of the publishers (in equilibrium) are then given by:
    
    \begin{gather}
        \begin{aligned}
            u_1(x^{eq}) & = r_1(x^{eq}) - \lambda d^0_1(x^{eq}_1) \\& \approx \frac{1}{2} \cdot \nu_1(x^{eq}) + \frac{1}{2} - 
            1 \cdot \frac{1}{2}\norm{(0.753, 0.326) - (0.950, 0.035)}^2 \\
            & \approx 0.415 - 0.062 = 0.353
        \end{aligned} \\
        \begin{aligned}
            u_2(x^{eq}) & = r_2(x^{eq}) - \lambda d^0_2(x^{eq}_2) \\& \approx \frac{1}{2} \cdot \nu_2(x^{eq}) + \frac{1}{2} - 
            1 \cdot \frac{1}{2}\norm{(0.741, 0.818) - (0.933, 0.773)}^2 \\
            & \approx 0.585 - 0.019 = 0.566
        \end{aligned}
    \end{gather}

\end{example}

The closed form of the PNE enables us to perform \emph{welfare analysis}. For example, deriving the publishers' welfare in equilibrium by $\lambda$ reveals that it has a global minimum in $\lambda = \frac{1}{n}$, that is, the point where the weight of $x^*$ and $x_0^i$ in the equilibrium strategy of player $i$ is equal. It decreases in $\lambda$ for $\lambda \leq \frac{1}{n}$ and increases for $\lambda \geq \frac{1}{n}$.
As for the users' welfare, under a regularity condition on $x_0$ and $x^*$, it decreases in $\lambda$.\footnote{
The regularity condition depends on the values \( d^i \coloneqq d(x_0^i, x^*) \) for each \( i \in N \), and its fulfillment is determined by the ratio between the mean and the variance of these distances.
In the case of independently and uniformly distributed $x_0$ and $x^*$, the regularity condition holds with a very high probability. When the regularity condition does not hold, the users' welfare has a unique extremum, which is a global minimum point.} Rigorous analysis is available within Appendix \ref{ape: lin analysis}.

We conclude this section by establishing that while applying a nonlinear function to the linear relative relevance might seem to generalize linear ranking functions, it turns out that this cannot be achieved.

\begin{theorem}
\label{theorem: linear only simplex}
    The only ranking functions of the form $r_i(x) = g\big(\nu _i (x) \big)$ for some differentiable $g:[-1, 1] \to \R$ are for linear \(g\).
\end{theorem}

\begin{proof}

     If $r$ is a valid ranking function, then in particular its image is contained in the simplex $\Delta^n$ for every publishers' game with $n=3,k=1,x^*=0,d(a,b)=(a-b)^2$. Thus for every such game, $\forall x\in X=[0,1]^3$ it holds that:
     \begin{equation}
         1 = \sum_{j=1}^n r_j(x) = \sum_{j=1}^n g(\nu_j(x))
     \end{equation}

    Define a function $f:X\to\mathbb{R}$ by $f(x)=\sum\limits_{j=1}^3 g(\nu_j(x))$. Since $f\equiv 1$ on $X$, 
    $\nabla f(x) = 
    \begin{pmatrix} 
        0 \\
        0 \\
        0 \\
    \end{pmatrix}$ 
for all $x\in X$. Note that in such games, $d^{*}(a) = (a-0)^2 = a^2$ and thus ${d^{*}}'(a) = 2a$.
So $\forall x\in X, \forall i \in [n]$ it holds:
\begin{equation}\label{eq: f derivative}
    0 = \frac{\partial f(x)}{\partial x_i}
 = \sum_{j\neq i} g'\big( \nu_j (x)\big) \cdot \frac{2x_i}{n-1} - g'\big ( \nu_i(x)\big ) \cdot 2 x_i
\end{equation}

Denote by $\psi : \mathbb{R}^n \to \mathbb{R} ^n $ the function defined by
\begin{equation}
    \psi_i(d) \coloneqq \frac{1}{n-1} \sum\limits_{j\neq i} d_j - d_i
\end{equation}
By \eqref{eq: f derivative}, $\forall i \in N$:
\begin{equation}
    0 = 2 x_i \cdot 
    \psi _i \biggl(  \Bigl (  g'\bigl(\nu_l(x)  \bigr ) \Bigr)_{l=1}^n \biggr) 
\end{equation}
Where $\Bigl (  g'\bigl(\nu_l(x)  \bigr ) \Bigr)_{l=1}^n$ stands for the vector 
$\begin{pmatrix}
    g'\bigl(\nu_1(x)  \bigr )
    &
    g'\bigl(\nu_2(x)  \bigr )
    & \dots &
    g'\bigl(\nu_n(x)  \bigr )
\end{pmatrix} \in \R^n$.

Hence $\forall x\in X, \forall i,j\in [n]$ the following holds:
\begin{equation}
    x_i \neq 0 \land x_j \neq 0 \implies 
    \psi _i \biggl(  \bigl (  g'\left(\nu_l(x)  \right ) \bigr)_{l=1}^n \biggr)
    = 0 =
    \psi _j \biggl(  \bigl (  g'\left(\nu_l(x)  \right ) \bigr)_{l=1}^n \biggr)
\end{equation}

We now use the following lemma (whose proof appears in Appendix \ref{ape: proofs}):

\begin{lemma}
\label{lemma: psi property}
    Let $\psi : \mathbb{R}^n \to \mathbb{R} ^n $ be the function defined by $\psi_i(d) = \frac{1}{n-1} \sum\limits_{j\neq i} d_j - d_i$.
    \\Then $\forall d\in \mathbb{R}^n, \forall i,j\in [n]$,
    \begin{equation}
        \psi_i(d) = \psi_j(d) \iff d_i = d_j
    \end{equation}
\end{lemma}

By Lemma \ref{lemma: psi property}, $\forall x\in X, \forall i,j\in [n]$ it holds that:
\begin{equation}\label{eq: useful eq}
    x_i \neq 0 \land x_j \neq 0 \implies 
    g'\left (v_i(x)\right ) = g'\left (v_j(x)\right )
\end{equation}

Now, we prove that $g$ must be linear by proving that $g'$ must be constant on $[-1,1]$. We do this in two steps: first, we show that $g'$ is constant on $(-0.75, 0.75)$ and then we show it is constant on the entire $[-1,1]$ as well.

For the first step, we look at the profiles $x_{\varepsilon} := \left (  \sqrt{0.5-\varepsilon}, \sqrt{0.5}, \sqrt{0.5+\varepsilon}  \right )$ for $\varepsilon \in (0,0.5)$. The linear relative relevance vector in this case is $\nu (x_\varepsilon) = (1.5\varepsilon, 0, -1.5\varepsilon)$. 
By applying \eqref{eq: useful eq} on $x_\varepsilon,i=1,j=2$ we get $g'(1.5\varepsilon)=g'(0)$.
By applying \eqref{eq: useful eq} on $x_\varepsilon,i=2,j=3$ 
we get $g'(-1.5\varepsilon)=g'(0)$.
So, we got $\forall \varepsilon \in (0,0.5), g'(-1.5 \varepsilon) = g'(0) = g'(1.5\varepsilon)$.
This means that $\forall t\in (0,0.75), g'(-t)=g'(0)=g'(t)$. 
Therefore $g'$ is constant on $(-0.75,0.75)$.

Now for the second step, let $t\in (-1,1)$ and we show that $g'(t)=g'(0)$.
Consider the profile $\hat{x}:=\left (  \sqrt{\frac{1-t}{2}}, \sqrt{\frac{1+t}{2}}, \sqrt{\frac{1+t}{2}}  \right )$.
The linear relative relevance vector in this case is $v(\hat{x}) = \left( t, -\frac{t}{2}, -\frac{t}{2} \right )$.
By applying \eqref{eq: useful eq} on $\hat{x},i=1,j=2$ we get $g'(t)=g' \left ( -\frac{t}{2}\right )$. Since $t\in (0,1)$, $-\frac{t}{2} \in (-0.75,0.75)$, so by the first step we have that $g' \left ( -\frac{t}{2}\right )=g(0)$. So $g'(t)=g'(0)$ as we wanted to show. This proves that $g'$ is constant on $(-1,1)$. Therefore, $g$ is linear on $(-1,1)$. Since $g$ is continuous, it is also linear on $[-1,1]$.
 
\end{proof}

\section{Empirical Results}
\label{sec: empirical}

In this section, we evaluate the performance of the PRP, the linear, and the softmax ranking functions in simulated environments that aim to reflect realistic better response dynamics. Empirical analysis of competitive retrieval settings is a difficult challenge, as discussed in \textcite{kurland2022competitive}. Conducting experiments using real data requires the implementation of each proposed ranking function into real systems, in which publishers repeatedly compete for exposure and receive feedback from a system that ranks according to the proposed approach. Therefore, we evaluate our ranking functions using a code-simulated environment which we call the \textbf{Discrete Better Response Dynamics Simulation}.\footnote{Code and results are available in \url{https://github.com/ireinman/The-Search-for-Stability}.} These dynamics are typical in real-world scenarios, where publishers endeavor to enhance the visibility and impressions of their content within the platform they operate. We focus on the canonical of $n=2$, meaning that two prominent publishers compete for search engine impressions.
See Appendix \ref{app: beyond n=2} for experiments and discussion beyond the two-publishers scenario.

\subsection{Simulation Details}\label{subsec: simulation details}

In the Discrete Better Response Dynamics Simulation, publishers start with their initial documents, and at each timestep, one publisher modifies their document to improve their utility. The modification is done by taking a step that maximizes the publisher's utility, from a predefined subset of possible directions $\mathcal{D}$ and step sizes $\mathcal{S}$. Formally, in each simulation, we consider a publishers' game $G$ with $n=2$, $k=2$,\footnote{The time complexity of each simulation timestep grows exponentially with $k$, therefore we used $k=2$. An analysis of the effect of $k$ on the results can be found within Appendix \ref{ape: k effect}.} $\lambda \in [0,1]$, $ d(x,y) = \frac{1}{k} ||x-y||_2^2 $ and uniformly drawn\footnote{Experiments with other distributions of \(x^*\) and \(x_0\) are reported in Appendix \ref{app: non iid dist}.} $x^* \in [0,1]^k$ and $x_0 \in [0,1]^{n \cdot k}$. A Discrete Better Response Dynamics Simulation is said to \textbf{converge} if it has reached an $\varepsilon$-PNE under the discreteness constraints (the constraints of predefined directions and step sizes) within no more than $T$ iterations.

To evaluate the performance of the different ranking functions, we have estimated the \emph{convergence ratio}--the proportion of simulations that converged--and the expected long-term publishers' and users' welfare. In case the simulation converged, we estimate the welfare measures over the long term using the welfare measures of the profile the simulation converged to. Otherwise, we average the welfare measures over the last $M$ rounds of the simulation, where $M$ is an additional simulation parameter. 
A justification for this estimation method is the empirically observed periodic nature of the dynamics that did not converge (see \S\ref{subsec: learning dynamics} for a detailed discussion). For each value of $\lambda$, we performed $500$ simulations and constructed a bootstrap confidence interval with a confidence level of 95\%. Algorithm \ref{alg: empirical code} provides the pseudo-code for a single simulation.\footnote{
We used $\mathcal{S} = \{2^{-6}, 2^{-5}, ..., 2^{-2}, 0.5, 0.6, ..., 1 \}$, $\;\mathcal{D} = \big \{ \frac{d}{||d||_2}: d \in \{-1,0,1\}^k \setminus \{ \vec{0} \} \big \}$,
$T=1000, \; M=900, \; \varepsilon=10^{-6}$, and a bootstrap sample size of $B = 500$.
}

\begin{algorithm}[tb]
\caption{Discrete Better Response Dynamics Simulation}\label{alg: empirical code}
\begin{algorithmic}[1]

\STATE {\bfseries Input:} $G, \mathcal{S}, \mathcal{D}, T, M, \epsilon$
\STATE $x^{(0)} \gets x_0$
\STATE $conv \gets False$
\FOR{$t=1,2,...T$} 
    \STATE $ \mathcal{I} \gets \text{the set of players that can improve their utility}$
    \STATE $\text{ by more than } \varepsilon \text{ by taking a step in direction } d\in\mathcal{D}$
    \STATE $\text{ with step size } s\in\mathcal{S}$
    \IF{ $\mathcal{I} \neq \emptyset$}
        \STATE Draw $i \in \mathcal{I}$ uniformly
        \STATE ${BR}_i \gets \underset{{(s,d) \in \mathcal{S} \times \mathcal{D}}}{\argmax} \Bigl \{u_i \bigl(x^{(t-1)}_i + s \cdot d, \; x^{(t-1)}_{-i} \bigr)  \Bigr \}$
        \STATE Draw $(s', d') \in {BR}_i$ uniformly
        \STATE $x^{(t)} \gets \bigl (x_i^{(t-1)} + s' \cdot d', \; x_{-i}^{(t-1)} \bigr)$
    \ELSE
        \STATE $conv \gets True$
        \STATE \textbf{return} {$\bigg(\mathcal{U}(x^{(t-1)}), \mathcal{V}(x^{(t-1)}), conv \bigg )$} 
    \ENDIF
    
\ENDFOR
\STATE $T_s \gets T - M + 1$
\STATE \textbf{return}{$ \biggr(
\text{Avg} \bigl \{ \mathcal{U} (x^{(t)}) \bigr \}_{t=T_s}^T, 
\text{Avg} \bigl \{ \mathcal{V} (x^{(t)}) \bigr \}_{t=T_s}^T,
conv 
\biggl)$}
    
\end{algorithmic}
\end{algorithm}

    


\subsection{Comparison of the PRP, Softmax, and Linear Ranking Functions}
\label{subsec: simulation results - between rankers}

\begin{figure}[tb]
\centering
\includegraphics[width=\textwidth,height=\textheight,keepaspectratio]{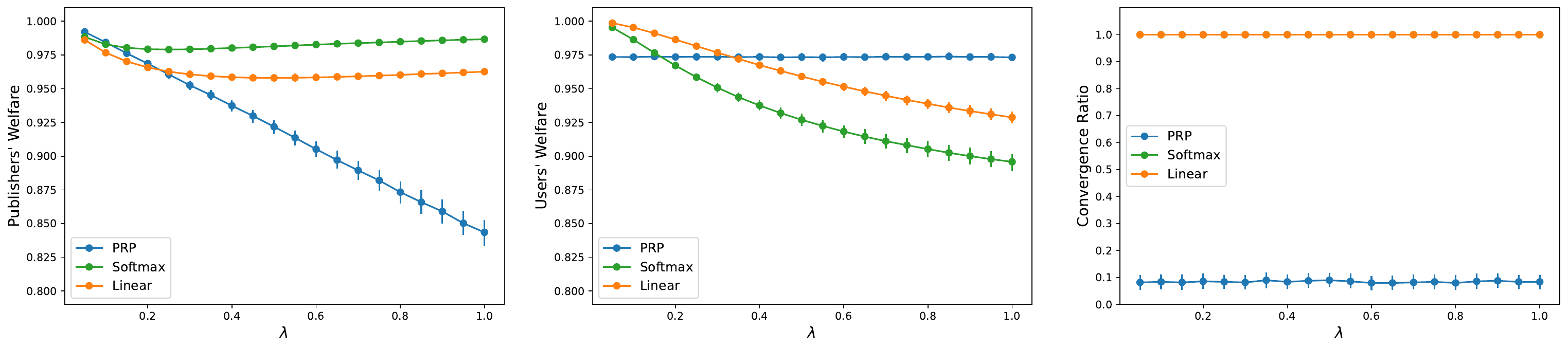}
\caption{Comparison of the PRP, softmax ($\beta = 1$) and linear ($a=\frac{1}{n}=0.5$) ranking functions. The linear and the softmax ranking functions had a convergence ratio of $1$.}
\Description[Comparison of the PRP, linear, and softmax ranking functions.]{The publishers' welfare, the users' welfare, and the convergence ratio are presented as a function of lambda for each of the three ranking functions. The convergence ratio for the PRP ranking function is approximately 0.1 for all lambda values.}
\label{fig: PRP_RRP}
\end{figure}

Figure \ref{fig: PRP_RRP} provides a comparative analysis of the three ranking functions: PRP, softmax (with $\beta = 1$), and linear (with $a=\frac{1}{n}$). 
The rightmost chart in Figure \ref{fig: PRP_RRP} highlights a fundamental shortcoming of the PRP ranking - most of the dynamics it induces do not converge. However, dynamics under the other two ranking functions always converge. This is a key advantage of these functions. A noteworthy point that emerged from the simulations but is not shown in the graphs is that convergence to an $\varepsilon$-PNE under the softmax and linear ranking functions was very rapid, typically occurring after only a few timesteps.

Another insight from Figure \ref{fig: PRP_RRP} is that the users' welfare of the PRP is roughly constant across $\lambda$ values. 
A plausible explanation is that in the case of PRP and the tested $\lambda$ range, $\lambda$ has little, if any, impact on the behavior of publishers. This conjecture might also explain why the publishers' welfare of the PRP appears to linearly decrease with $\lambda$: the dynamics remain the same, but the term $ - \lambda \cdot \sumi d^0_i(x_i)$ in the publishers' welfare gains more influence as $\lambda$ increases. Note that the PRP ranking function appears to sacrifice stability in pursuit of increased users' welfare.

In the linear ranking function, the effect of $\lambda$ on the welfare measures matches the theoretical results presented in \S\ref{linear_theory}--the publishers' welfare obtains a minimum in $\lambda = \frac{1}{n}$,
and the users' welfare is monotonically decreasing in $\lambda$.\footnote{The likelihood of the regularity condition mentioned in \S\ref{linear_theory} under uniform distribution of $x_0$ and $x^*$ is very close to $1$. This explains why the results shown in Figure \ref{fig: PRP_RRP} correlate with the theoretical result for when the condition does hold.} 
Another observation from Figure \ref{fig: PRP_RRP} is that the softmax ranking function exhibits similar trends. In this figure, an inherent trade-off in our model can be seen - an increase in publishers' welfare leads to a decrease in users' welfare and vice versa.

It seems that the softmax ranking function has higher publishers' welfare and lower users' welfare than the linear ranking function.
Yet, tuning the inverse temperature constant $\beta$ or the slope $a$ can change this.
We now provide an empirical analysis of these hyperparameters' influence.

\subsection{The Effect of Hyperparameters Selection}
\label{subsec: simulation results - within rankers}

\begin{figure}[tb]
\centering
\includegraphics[width=\textwidth,height=\textheight,keepaspectratio]{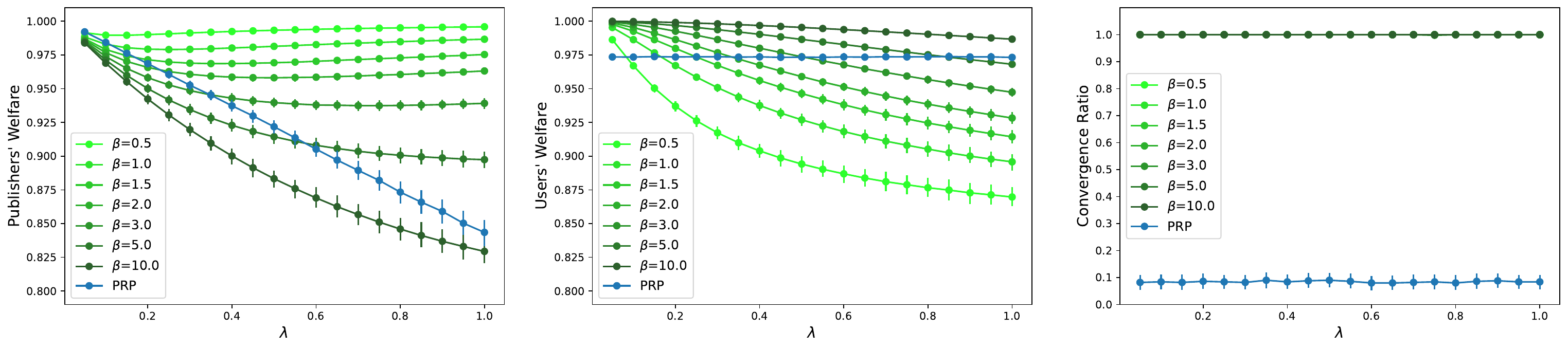}
\caption{The effect of $\beta$ on the welfare measures and the convergence ratio in the softmax ranking function.
All softmax ranking functions had a convergence ratio of $1$.}
\label{fig: softmax_beta}
\Description[The effect of the inverse temperature constant beta]{Figure 2 presents the welfare measure and the convergence ratio as a function of lambda for different values of beta in the softmax ranking function, as well as for the PRP ranking function. A larger beta improves users' welfare but worsens publishers' welfare. Importantly, some values of beta have better users' welfare than the PRP.}
\end{figure}

Figure \ref{fig: softmax_beta} provides a comparative analysis of the influence of the inverse temperature constant $\beta$ in the softmax ranking function. It is evident that $\beta$ serves as a means to control the inherent trade-off between publishers' welfare and users' welfare. A search engine designer can tune $\beta$ according to the relative importance of the two welfares within her domain and according to her estimation of $\lambda$. Importantly, the convergence ratio of $\tilde{r}^{\beta}$ is $1$ for all tested values of $\beta$. Therefore, the softmax with high $\beta$ values (for example, $\beta \in \{5, 10 \}$) manages to surpass the PRP in terms of users' welfare while maintaining the desired property of $100\%$ convergence. The cost is sacrificing publishers' welfare. Therefore, we claim that the softmax ranking functions offer viable alternatives to the PRP function.

As mentioned in \S\ref{sec: ranking functions}, the PRP ranking function maximizes the users' welfare for a fixed profile. The reason why softmax ranking functions with high $\beta$ values nevertheless manage to achieve higher users' welfare than the PRP is that the PRP is only short-term optimal, 
while the softmax ranking functions preserve high users' welfare in the long term as well.
Unlike the softmax ranking functions, most of the dynamics the PRP induces do not converge. In such ecosystems, the publishers repeatedly reach profiles with sub-optimal users' welfare, and this harms the long-term users' welfare of the PRP. This emphasizes the need for the search for stability.

\begin{figure}[tb]
\centering
\includegraphics[width=\textwidth,height=\textheight,keepaspectratio]{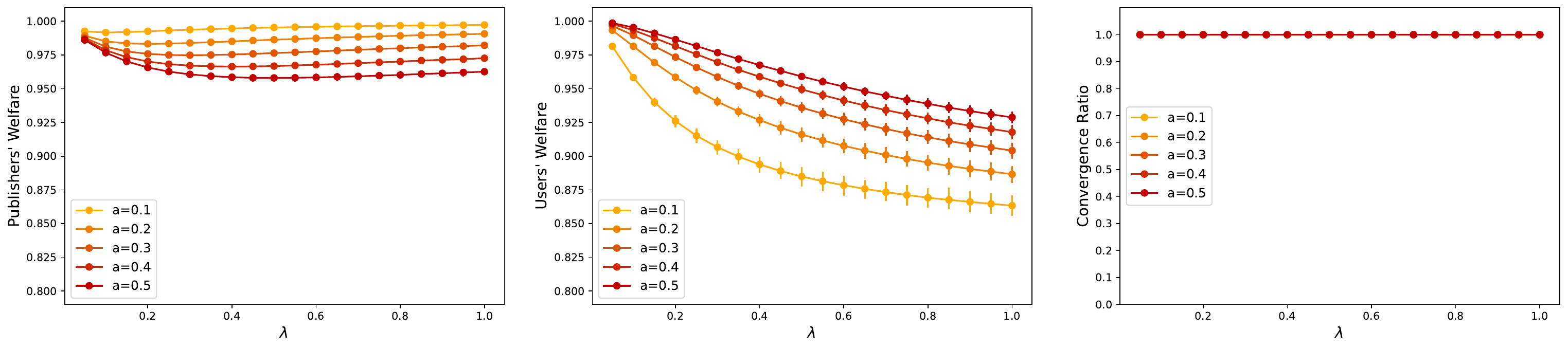}
\caption{The effect of $a$ on the welfare measures and the convergence ratio in the linear ranking function. All linear ranking functions had a convergence ratio of $1$.}
\label{fig: linear_slope}
\Description[The effect of the slope a of the linear ranking functions]{Figure 3 presents the welfare measure and the convergence ratio as a function of lambda for different values of a in the linear ranking function. The larger a, the better the users' welfare and the worse the publishers' welfare.}
\end{figure}

Now, let us delve deeper into the linear ranking function and compare it to the softmax. Figure \ref{fig: linear_slope} provides a comparative analysis of the influence of the slope $a$ in the linear ranking function. As was the case for the softmax ranking function, $\hat{r}^{a}$ has a convergence ratio of $1$ for all tested $a$ values. This is not surprising, as we proved it in Corollary \ref{corollary: linear con}.

Just like $\beta$ in the softmax functions, the slope $a$ allows the search engine designer to control the publisher-user trade-off. There is, however, a core difference between the tuning of $a$ and $\beta$. While the range of $a$ values for which $\hat{r}^{a}$ is a valid ranking function is bounded from above by $\frac{1}{n}$ (see \S\ref{linear_theory}), one can set $\beta$ to be arbitrarily large. This means that using the softmax ranking function it is possible to achieve high users' welfare values that cannot be obtained using the linear ranking function. Nevertheless, the linear still has a significant advantage - it induces a potential game. 
As shown in \S\ref{linear_theory}, the existence of a potential function enables rigorous theoretical analysis and guarantees the convergence of $\varepsilon$-better response dynamics.

\subsection{Learning Dynamics Under the PRP Ranking Function}\label{subsec: learning dynamics}

In this section, we delve into the learning dynamics of games induced by the PRP ranking function. While the dynamics in games induced by the softmax and the linear ranking functions are relatively easy to anticipate, as they converge to an $\varepsilon$-PNE in a few timesteps, PRP-based games often exhibit completely different phenomena. We will explore these intriguing dynamics in detail.

\begin{figure}
\centering
\includegraphics[width=.4\textwidth,height=\textheight,keepaspectratio]{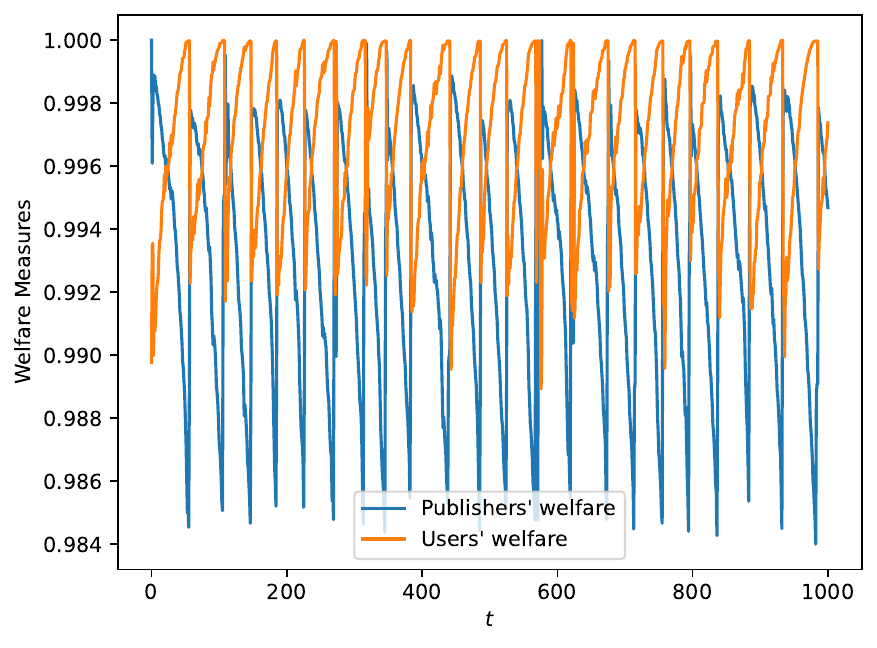}
\caption{The welfare measures during a Discrete Better Response Dynamics Simulation with the PRP ranking function on a representative instance of a publishers' game with $\lambda = 0.5, n=2 \text{ and } k=2$. The figure illustrates the cyclic nature of the dynamics over time.}
\label{fig: prp welfares periodicity}
\Description[A periodic behavior in PRP games]{Figure 4 presents the users' welfare and the publishers' welfare as a function of the iteration t in a representative PRP game. Both plots look like periodic functions.}
\end{figure}

During the examination of the PRP ranking function, we noticed that the vast majority (about $99.5\%$) of the dynamics it induces belong to one of two types of dynamics.\footnote{In the rare dynamics that do not fit into either of the two main types of dynamics, the simulation converges after more than one timestep. The convergence in those cases can be attributed to the fact that the step sizes are bounded from below.} One type, which constitutes a minority of the cases, is when the simulation converges already in the first round because no publisher can improve their utility under the constraints of the discrete simulation. In those cases, the publisher whose initial document is farther from $x^*$ (the information need) cannot get closer to $x^*$ than the other publisher in a single step, and therefore cannot improve her utility in a single step. Moreover, the other publisher gets the maximum utility, $1$, because she is the closest to $x^*$ and she is at her initial document, therefore she also cannot improve her utility in a single step. 
In the other type of simulations, which constitutes most of the cases, the publishers enter a pseudo-periodic\footnote{We use the term ``pseudo-periodic'' instead of ``periodic'' since the publisher who deviates in each timestep is chosen at random from all the publishers who can improve their utility by more than $\varepsilon$. Because of this randomness, the dynamics that exhibit a periodic nature usually display variations between cycles, thus justifying the term ``pseudo-periodic''.} behavior that never converges. Figure \ref{fig: prp welfares periodicity} presents the welfare measures during one representative instance of the latter type of simulations.

The periodic nature of the publishers' behavior is evident in Figure \ref{fig: prp welfares periodicity}. In each cycle, there is a period where the publishers compete to provide the closest document to $x^*$. Each of them repeatedly changes her document (strategy) by the smallest margin sufficient for her to get closer to the information need than her competitor. This competition, which is accompanied by an increase in the users' welfare and a decrease in the publishers' welfare, continues until one of the players resigns and abruptly returns as close as she can to her initial document. Her decision to resign is either because the discrete nature of the simulation prevents her from bypassing the other publisher, or because further bypassing the other publisher is not beneficial for her since it positions her too far from her initial document.
Immediately after this resignation, the publisher who won the competition takes a big step toward her initial document while still maintaining her position as the closest to $x^*$, since as long as she is still closer to $x^*$ than the other publisher, getting closer to her initial document is trivially beneficial for her. This mutual withdrawal is accompanied by a sharp decrease in users' welfare and a sharp increase in publishers' welfare. When the losing publisher notices the winning publisher's retreat from $x^*$, she exploits the situation to get back into the competition and bypass her rival. The entire process is then repeated over and over.

It can be seen that the PRP reaches profiles with users' welfare of almost $1$, which is the optimal users' welfare value, but it does not stay there. This explains why, despite the PRP being optimized for short-term relevance (that is, for a fixed profile, it maximizes the users' welfare), its instability harms its long-term performance. This phenomenon once again emphasizes the need for the search for stability.

The periodicity is best observed when animating the documents in a two-dimensional embedding space, which can be done using the code provided in the supplementary material. An example video is provided as well.\footnote{The video can be downloaded from our GitHub repository: \url{https://github.com/ireinman/The-Search-for-Stability/blob/main/figures/animation_example.mp4}.} In this visualization, one can notice that during the simulation, each publisher stays close to the line segment that connects $x^*$ with her initial document. This phenomenon is correlated with the theoretical result of Lemma \ref{lemma: the line dominates} (in Appendix \ref{ape: euclidean norm}), which determines that in games induced by various ranking functions, including the PRP, any strategy that is outside this line segment is strictly dominated.\footnote{The fact that during the simulation players do not play exactly on the line segment but rather just close to it is due to the discrete nature of our simulation.}

\section{Discussion}
\label{sec: discussion}

We studied a game-theoretic model for information retrieval, which models documents as vectors in a continuous space, accounts for original content, and distinguishes between publishers' and users' welfare. We have provided both theoretical and empirical negative results about the stability of ecosystems in which the PRP is used to rank the documents in the corpus and demonstrated how instability harms long-term users' welfare. Driven by the search for stability, we propose two alternative ranking function families - the linear and the softmax ranking function families. We focus on a strong criterion for stability, which is the convergence of better response dynamics. 
For the linear ranking function, we have theoretically established the convergence of $\varepsilon$-better response dynamics, and we have empirically demonstrated that the softmax ranking function fosters stable ecosystems.
The convergence of any $\varepsilon$-better response dynamics in games induced by the linear ranking function is a critical finding, as such dynamics mirror the competitive landscape observed in the real world among publishers. Moreover, we empirically showed a trade-off between publishers' welfare and users' welfare and provided means for a search engine designer to control it.

\paragraph{Solution concept and stability guarantee}
A natural question in the context of our analysis is regarding the consideration of a \emph{pure} Nash equilibrium existence as a desirable property. One may claim that the existence of a \emph{mixed} Nash equilibrium, in which publishers are allowed to engage in mixed strategies (that is, randomize pure actions), is sufficient as a stability criterion of the ecosystem induced by the ranking scheme. We provide several justifications for our rather stronger requirement.

First, while it may be plausible that publishers engage in mixed strategies (for instance, by using generative AI for content creation), a mixed Nash equilibrium results in unpredictable outcomes for the designer. In contrast, a PNE reflects a scenario where all publishers play deterministically, leading to a more stable ecosystem. Predicting and monitoring the content available within the platform becomes easier when publishers converge to a PNE.
Second, we highlight that when publishers engage with pure strategies, users can be sure they will always find relevant content on the platform. Indeed, under some different modeling assumptions, generating content deterministically may be less desirable (for example, users may become bored if content is generated deterministically). In some contexts, this is not the case. For instance, whenever \(x^*\) is a piece of information that the user expects to find in a search engine, she always wants to find the most informative document, and boredom does not play a role. This stands in contrast to other use cases, such as content consumed for entertainment purposes.
Another reason to prefer PNE is the effort required for generating randomized content. For instance, randomization can be done using LLMs, which is expensive when used at scale. See also the discussion in \textcite{ben2018game} on PNE as a property indicating stability.

Importantly, much of our analysis focuses not only on the existence of a PNE but also on a stronger stability property: the guarantee that any better-response dynamics will converge. This property is particularly powerful in real-world applications, where the specific structure of content modifications—such as how changes in the text space translate to the embedding space—is often unknown to the designer. Anticipating these transformations is challenging, but under the mild assumption that publishers can modify documents to improve their utility (as determined by the embedding space), strategic content modifications are guaranteed to converge, even without the designer explicitly understanding their precise structure.

\paragraph{Limitations and future work}
We find it important to discuss several limitations of this work. 
First, our model assumes a specific form of the publishers' utilities (winning probability minus integrity cost). Despite being restrictive, the assumption is quite standard, as the two terms usually do not interact.
We assume publishers' full knowledge of the ranking function and the information need. This is a very common assumption when publishers compete in a \emph{niche domain} – such publishers are usually experts who are familiar with the requirements and needs of users in their domain, as well as with the platform's algorithm.
Following the literature on information retrieval games, our model assumes that publishers only gain utility from being ranked first.
Additionally, we experimented with a specific better-response dynamics structure and specific distance functions. Importantly, for cases where any better response dynamics is guaranteed to converge to a unique equilibrium, these specifications are not restrictive at all.
Another limitation in the experimental setting is the restriction of uniformly i.i.d. distributing $x_0$ and $x^*$. In Appendix \ref{app: non iid dist} we provide preliminary results beyond this case, in which all trends agree with the ones presented in the paper.

Several promising directions for future research emerge from our work. First, extending the model to incorporate heterogeneous cost parameters (that is, assigning each publisher an individual $\lambda_i$ rather than assuming a common $\lambda$) would better capture the asymmetries found in real-world publisher behavior. We note that many of our results and techniques naturally extend to this more general setting.

Second, a more refined analysis of the \emph{rate of convergence} to equilibrium under various ranking functions is of both theoretical and practical interest. In real-world systems, designers often care about the speed at which a stable state is reached. Furthermore, while some information needs are timeless, others are time-sensitive, making quick convergence essential for maintaining user relevance (for instance, consider news or trending topics). Modeling such \emph{time-dependent preferences} among publishers and users presents a valuable avenue for future work.

A related extension is to study \emph{non-stationary information needs}, that is, scenarios in which the target query itself evolves over time, potentially as a function of previously published documents. This captures the well-documented phenomenon of \emph{preference shifts} in recommendation systems \parencite{dean2022preference}. Formally, such a setting could be modeled using the framework of stochastic games, and we leave its integration into the publishers' game model as an interesting open direction.

Finally, while our model reflects considerations such as \emph{integrity toward a publisher’s preferred ideal content} (as captured by distance from the initial document $x^0_i$), an alternative approach would define cost in terms of \emph{modification effort}---that is, the distance from the previously published document. This alternative cost model may better reflect scenarios where the dominant constraint is the \emph{effort or friction} involved in repeatedly editing or rewriting content.

\begin{acks}
  This work was supported by funding from the European Research Council (ERC) under the European Union’s Horizon 2020 research and innovation programme (grant agreement 740435).
  We would like to thank the associate editor and the anonymous reviewers for their helpful comments.
  We also thank Yotam Gafni for his helpful comments on an early version of the paper. All authors contributed equally to this work.
\end{acks}

\printbibliography

\appendix\label{ape}

\section{Equilibrium Analysis with the Euclidean Norm}
\label{ape: euclidean norm}

In this section, we provide an equilibrium analysis of publishers' games that are based on the Euclidean norm, that is, publishers' games with $d(x,y)=\frac{1}{\sqrt{k}} \norm{x-y}_2$. For each player $i \in N$, we define $L_i$ as the line segment that connects the points $ x_0^i $ (her initial document) and $x^* $ (the information need). Formally, $L_i$  is defined as $ L_i \coloneqq \{t x^* + (1 - t)x_0^i : t \in [0,1]\}$. We use the notation $P_{L_i}(x_i)$ for the projection of $x_i$ on $L_i$. We will make use of the following definitions:

\begin{definition}
    We say that a strategy $x_i \in X_i$ \textbf{strictly dominates} strategy $x'_i \in X_i$ (or: $x'_i$ is \textbf{strictly dominated by} $x_i$) if for every strategy profile of the other publishers $x_{-i}\in X_{-i}$, $x_i$ is a better response than $x'_i$ with respect to $x_{-i}$. \\
    A strategy $x_i\in X_i$ is called a \textbf{strictly dominant strategy} if it dominates any other strategy of player $i$.\\
    A strategy $x_i\in X_i$ is called a \textbf{strictly dominated strategy} if it is strictly dominated by another strategy of player $i$.
\end{definition}

Note that a strictly dominated strategy is never played in a pure Nash equilibrium. 
Interestingly, in many publishers' games, a significant number of strategies are strictly dominated.

\begin{lemma}
\label{lemma: the line dominates}
    Let $d$ be a function of the form $d(a,b)=\frac{\norm{a-b}^p}{C}$ for some $C>0$ and $p\geq 1$, where $\norm{\cdot}$ is the Euclidean ($L_2$) norm.
    Let $G$ be a publishers' game with $d$ as its semi-metric and whose ranking function \(r\) satisfies that \(r_i(x)\) is decreasing (not necessarily strictly) in \(d^*(x_i)\). Then, for every player $i\in N$, any strategy $x_i \notin L_i $ is strictly dominated by $P_{L_i}(x_i)$.
\end{lemma}
\begin{proof}
        We use the fact that since $L_i$ is convex and closed,
        $P_{L_i}(x_i)$ is well defined and unique, and 
        the following holds \cite{bertsekas1997nonlinear}:
        \begin{equation}
            \label{eq:projection}
            \forall y \in L_i,
            \left \langle 
                x_i - P_{L_i}(x_i), y - P_{L_i}(x_i)
            \right \rangle
            \leq 0
        \end{equation}

    where $\langle \cdot, \cdot \rangle$ stands for the standard inner product. 
    
    Also, if $x_i \notin L_i$, then $P_{L_i}(x_i) \neq x_i$ 
    so $\norm{x_i - P_{L_i}(x_i)} > 0$. We get that for every 
    $y \in L_i $:
    \begin{equation}
        \begin{aligned}
        \norm{x_i - y} ^2 &= \norm{x_i - P_{L_i}(x_i) + P_{L_i}(x_i) - y}^2 \\
    &= \norm{x_i - P_{L_i}(x_i)}^2 + \norm{P_{L_i}(x_i) - y}^2
    + 2 \left \langle x_i - P_{L_i}(x_i), P_{L_i}(x_i) - y \right \rangle \\
    & \underset{\eqref{eq:projection}}{\geq} \norm{x_i - P_{L_i}(x_i)}^2 + \norm{P_{L_i}(x_i) - y}^2
    > \norm{P_{L_i}(x_i) - y}^2 
    \end{aligned}
    \end{equation}

    Hence in particular $\norm{x_i - x^*} > \norm{P_{L_i}(x_i) - x^*}$, and thus $d^*(x_i) > d^* \left( P_{L_i}(x_i) \right) $,
    so by our assumption on \(r\) we have $r_i(x_i, x_{-i}) \leq r_i(P_{L_i}(x_i), x_{-i}) \; \forall 
    x_{-i} \in X_{-i}$. \\
    In the same way, $\norm{x_i - x_i^0} > \norm{P_{L_i}(x_i) - x_i^0}$,
    so $d^0_i (x_i) > d^0_i \left (P_{L_i}(x_i) \right )$.
    By combining these two inequalities we get that for any \(x_{-i} \in X_{-i}\):
    \begin{equation}
    \begin{aligned}
      u_i (x_i, x_{-i})  & = r_i (x_i, x_{-i}) - \lambda d^0_i(x_i)  \\ & < 
      r_i(P_{L_i}(x_i), x_{-i}) - \lambda d^0_i \left (P_{L_i}(x_i) \right )
       = u_i \left (P_{L_i}(x_i), x_{-i} \right )  
    \end{aligned}
    \end{equation}
    so by definition $x_i$ is strictly dominated by $P_{L_i}(x_i)$ in $G$.
\end{proof}

Note that the condition about the ranking function in Lemma \ref{lemma: the line dominates} is satisfied by all ranking functions discussed in the paper. The Lemma implies, in particular, that players would never play outside their line segment $L_i$ in a PNE. Lemma \ref{lemma: euclidean norm property}, unlike Lemma \ref{lemma: the line dominates}, does not hold for the squared Euclidean norm, and this is a main reason for the difference between the games induced by these two semi-metrics.

\begin{lemma}
\label{lemma: euclidean norm property}
    Let G be a publishers' game with $d(x,y)=\frac{1}{\sqrt{k}}||x-y||_2$. Then, for every player $ i \in N$, $ \forall x_i\in L_i$:
    \begin{equation}
        d^*(x_i) + d^0_i(x_i) = d(x_0^i, x^*)
    \end{equation}
\end{lemma}
\begin{proof}
Let G be a publishers' game with $d(x,y)=\frac{1}{\sqrt{k}}||x-y||_2$, and let $i \in N, \; x_i \in L_i$. By definition, $ \exists t \in [0, 1]$ s.t. $x_i = tx^* + (1 - t)x_0^i$.
\begin{equation}
    \begin{aligned}
    d(x_0^i, x^*) & = \frac{||x_0^i - x^*||_2}{\sqrt{k}} = 
    t \cdot \frac{||x^* - x_0^i||_2}{\sqrt{k}} + (1 - t) \cdot \frac{||x_0^i - x^*||_2}{\sqrt{k}} \\
    & = \frac{||tx^* - tx_0^i + x_0^i - x_0^i||_2}{\sqrt{k}} + 
    \frac{||(1 - t)x_0^i - (1 - t)x^*||_2}{\sqrt{k}} \\
    & = \frac{||tx^* + (1-t)x_0^i - x_0^i||_2}{\sqrt{k}} + 
    \frac{||tx^* + (1 - t)x_0^i - x^*||_2}{\sqrt{k}} \\
    & = \frac{||x_i - x_0^i||_2}{\sqrt{k}} + 
    \frac{||x_i - x^*||_2}{\sqrt{k}} = d^0_i(x_i) + d^*(x_i)
\end{aligned}
\end{equation}

\end{proof}

Note that the term $d(x_0^i, x^*)$ depends only on the game parameters and does not depend on the strategy player $i$ chooses to play. The above two Lemmas allow us to perform a thorough analysis of PNEs in games induced by the Euclidean norm and the linear ranking function.

\begin{lemma}
\label{lemma: euclidean pne}
    Let G be a publishers' game with $r = \hat{r}$ and $d(x,y)=\frac{||x-y||_2}{\sqrt{k}}$.
    \begin{itemize}
        \item If $\lan < 1$, then $x^{eq}$ defined by $x^{eq}_i = x^*, \; \forall i \in N$ is the only PNE in G, and $\forall i \in N, \; x^*$ is a strictly dominant strategy for player $i$.
        \item If $\lan = 1$, then $x^{eq}$ is a PNE in G if and only if $\forall i \in N, \; x^{eq}_i \in L_i$.
        \item If $\lan > 1$, then $x^{eq} = x_0$ is the only PNE in G, and $\forall i \in N, \; x_0^i$ is a strictly dominant strategy for player $i$.
    \end{itemize}
\end{lemma}

\begin{proof}
    Let G be a publishers' game with $r = \hat{r}$ and $d(x,y)=\frac{||x-y||_2}{\sqrt{k}}$. First, we notice that:
    \begin{equation}
        \begin{aligned}
    u_i(x) &= r_i(x) - \lambda d_i^0(x_i) = 
    \frac{\nu_i(x)}{n} + \frac{1}{n} - \lambda d_i^0(x_i) 
    = \bigl ( \sum\limits_{\substack{j \in N \setminus \{i\}}} \frac{d^*(x_j)}{n(n-1)} \bigr ) - \frac{d^*(x_i)}{n} + \frac{1}{n} - \lambda d_i^0(x_i)
\end{aligned}
\end{equation}

Now, for every player $i\in N$ and $\forall x_{-i} \in X_{-i}$:
\begin{equation}
    \begin{aligned}
    \argmax_{x_i \in X_i}\{u_i(x_i, x_{-i})\} &
     \overset{(1)}{{=}} \argmax_{x_i \in L_i}\{u_i(x_i, x_{-i})\} 
     \overset{(2)}{{=}} \argmax_{x_i \in L_i}\Bigl\{ - \frac{d^*(x_i)}{n}  - \lambda d_0(x_i) \Bigr\} \\
     & = \argmin_{x_i \in L_i}\Bigl\{ d^*(x_i)  + \lan d_0(x_i) \Bigr\}\overset{(3)}{{=}}
     \argmin_{x_i \in L_i}\Bigl\{ d(x_0^i, x^*) - d_i^0(x_i)  + \lan d_0(x_i) \Bigr\} \\
     &=
     \argmin_{x_i \in L_i}\Bigl\{  ( \lan - 1) d_i^0(x_i) \Bigr\}
     = 
     \begin{cases}
        \{x^*\}, \text{ if } \lan < 1 \\
        L_i, \text{ if } \lan = 1 \\
        \{x_0^i\}, \text{ if } \lan > 1
    \end{cases}   
\end{aligned}
\end{equation}

When transition $(1)$ is based on Lemma \ref{lemma: the line dominates}, transition $(2)$ is substitution and removal of terms that do not depend on $x_i$ and transition $(3)$ is based on Lemma \ref{lemma: euclidean norm property}.
This completes the proof, since a profile $x$ is a PNE if and only if for every player $i\in N$ it holds that $x_i \in \argmax\limits_{x'_i \in X_i}\{u_i(x'_i, x_{-i})\}$.

\end{proof}

The resulting PNEs are degenerate in the sense that small changes in the value of $\lambda$ or $n$ can cause a complete change in both the PNEs and the publishers' dominant strategies.

\section{The Linear Ranking Function: Welfare Measures in Equilibrium}
\label{ape: lin analysis}

Let $G$ be a publishers' game with $r = \hat{r}$ (that is, the linear ranking function with slope $a = \frac{1}{n}$) and $d(x,y)=\frac{1}{k}||x-y||^2_2$.
According to Lemma \ref{lemma: linear PNE}, G has exactly one PNE which is $x^{eq}_i = \frac{x^* + \lan x_0^i}{1 + \lambda n}$. We aim to analyze the effect of the game parameters on $\mathcal{U}(x^{eq})$, the publishers' welfare in equilibrium, and on $\mathcal{V}(x^{eq})$, the users' welfare in equilibrium. We focus our analysis on the effect of $\lambda$, but a similar analysis can be made for other game parameters. Let us find a closed form for $\mathcal{U}(x^{eq})$ and for $\mathcal{V}(x^{eq})$. To this end, we denote by $d^i \coloneqq d(x_0^i, x^*)$ the distance of publisher $i$'s initial document from the information need. We also denote by $d^{-i} \coloneqq \frac{1}{n-1} \Bigl( \sum_{j \in N \setminus \{i\}} d^j \Bigr)$ the average of this distances for all publishers but $i$ and by 
$\overline{d} \coloneqq \frac{1}{n} \sum_{i=1}^n d^i$ the average of those distances for all publishers. Lastly, let us denote $d' \coloneqq \frac{1}{n}\sum_{i=1}^n d^i \cdot (d^{-i} - d^i)$. Let us now perform some simple calculations and substitutions. Firstly, the distances of player $i$'s equilibrium strategy from the information need and from her initial document are given by:
\begin{equation}
\begin{aligned}
   d^*(x^{eq}_i) & = \frac{1}{k} \norm{x^{eq}_i - x^*}_2^2
   = \frac{1}{k} \norm{\frac{x^* + \lan x_0^i}{1 + \lan} - x^*}_2^2 
   \\ & = \frac{(\lan)^2}{k(1+\lan)^2} \norm{x_0^i - x^*}_2^2 = 
   \frac{(\lan)^2 }{(1+\lan)^2} d^i
   \end{aligned}
\end{equation}

\begin{equation}
\begin{aligned}
   d^0_i(x^{eq}_i) & = \frac{1}{k} \norm{x^{eq}_i - x_0^i}_2^2 = \frac{1}{k} \norm{\frac{x^* + \lan x_0^i}{1 + \lan} - x_0^i}_2^2 
   \\& = \frac{1}{k(1+\lan)^2} \norm{x^* - x_0^i}_2^2= 
   \frac{1}{(1+\lan)^2} d^i
   \end{aligned}
\end{equation}

Player $i$'s linear relative relevance and player $i$'s probability to be ranked first are given by the following terms, respectively:
\begin{equation}
\begin{aligned}
        \nu_i(x^{eq}) & = \frac{1}{n-1} \Bigl( \sum_{j \in N \setminus \{i\}} d^*(x^{eq}_j) \Bigr) - d^*(x^{eq}_i) 
        \\ & = \frac{1}{n-1} \Bigl( \sum_{j \in N \setminus \{i\}} \frac{(\lan)^2 d^j}{(1+\lan)^2} \Bigr) - \frac{(\lan)^2 d^i}{(1+\lan)^2} 
        \\ & =
        \frac{(\lan)^2}{(1+\lan)^2} (d^{-i} - d^i)
\end{aligned}
\end{equation}
\begin{equation}
        \hat{r}_i(x^{eq}) = \frac{1}{n} \nu_i(x^{eq}) + \frac{1}{n} = \frac{\lambda ^ 2 n}{(1+\lan)^2} (d^{-i} - d^i) + \frac{1}{n}
\end{equation}

Now we can substitute what we found to find the equilibrium utility of player $i$:
\begin{equation}
\begin{aligned}
        u_i(x^{eq}) &= \hat{r}_i(x^{eq}) - \lambda d^0_i(x^{eq}_i)
        \\ &= \frac{\lambda ^ 2 n}{(1+\lan)^2} (d^{-i} - d^i) + \frac{1}{n} - \frac{\lambda d^i}{(1+\lan)^2}
        \\ &= \frac{\lambda ^ 2 n (d^{-i} - d^i) - \lambda d^i}{(1+\lan)^2} + \frac{1}{n}
\end{aligned}
\end{equation}

We are now ready to substitute the previous equations in order to find a closed form for the welfare measures in equilibrium. The publishers' welfare in equilibrium is simply:
\begin{equation}
        \mathcal{U}(x^{eq}) = 1 - \lambda \sum_{i=1}^n d^0_i(x^{eq}_i)  = 1 - \lambda \sum_{i=1}^n \frac{1}{(1+\lan)^2} d^i 
        = 1 - \frac{\lan}{(1+\lan)^2} \overline{d} 
\end{equation}
and the users' welfare is given by:
\begin{equation}
    \begin{aligned}
    \mathcal{V}(x^{eq})  & = 1 - \sum_{i=1}^n \hat{r}_i(x^{eq}) \cdot d^*(x^{eq}_i) = 
    1 - \sum_{i=1}^n \Bigl( \frac{\lambda ^ 2 n}{(1+\lan)^2} (d^{-i} - d^i) + \frac{1}{n} \Bigr) \cdot \frac{(\lan)^2 }{(1+\lan)^2} d^i \\
    & = 1 - \frac{(\lan)^2}{(1+\lan)^2} \cdot \frac{1}{n} \sum_{i=1}^n d^i - \frac{(\lan)^4}{(1+\lan)^4} \cdot \frac{1}{n} \sum_{i=1}^n d^i \cdot (d^{-i} - d^i) \\ 
    & = 1 - \frac{(\lan)^2}{(1+\lan)^2} \cdot \overline{d} - \frac{(\lan)^4}{(1+\lan)^4} \cdot d'
    \end{aligned}
\end{equation}

We will now examine the impact of $\lambda$ on both publishers' and users' welfare in equilibrium. This parameter represents the publishers' incentive to adhere to their initial document. Therefore, understanding the impact of $\lambda$ is vital for system designers striving to optimize welfare measures.

\subsection{Publishers' Welfare Analysis}

We start by finding the partial derivative of $\mathcal{U}(x^{eq})$ with respect to $\lambda$.

\begin{equation}
    \frac{\partial \mathcal{U}(x^{eq})}{\partial \lambda}
    = - \frac{n(1+\lan)^2 - 2\lambda n^2(1+\lan)}{(1+\lan)^4}\cdot \overline{d}
    = \frac{(1+\lan)- 2 \lambda n}{(1+\lan)^3} \cdot \overline{d}n
    = \frac{1 - \lan}{(1+\lan)^3} \cdot \overline{d} n
\end{equation}

The denominator is always positive, therefore $\frac{\partial \mathcal{U}(x^{eq})}{\partial \lambda}$ is positive when $\lambda < \frac{1}{n}$, zero when $\lambda = \frac{1}{n}$ and negative when $\lambda > \frac{1}{n}$. We get that  $\lambda = \frac{1}{n}$ is a global minimum point of $\mathcal{U}(x^{eq})$.

\subsection{Users' Welfare Analysis}

We start by finding the partial derivative of $\mathcal{V}(x^{eq})$ with respect to $\lambda$. Recall that the users' welfare in equilibrium is given by:

\begin{equation}
    \mathcal{V}(x^{eq}) = 1 - \frac{(\lan)^2}{(1+\lan)^2} \cdot \overline{d} - \frac{(\lan)^4}{(1+\lan)^4} \cdot d'
\end{equation}

For ease of derivation, let us denote $f(t) \coloneqq 1 - t \overline{d} - t^2 d'$ and $g(\lambda) \coloneqq \frac{(\lan)^2}{(1+\lan)^2}$. Let us write the derivatives of these two functions: $f'(t) = - \overline{d} - 2 t d'$, $g'(\lambda) = \frac{2 \lambda n^2}{(1 + \lan)^3}$. We can now write:

\begin{equation}
    \mathcal{V}(x^{eq}) = f(g(\lambda))
\end{equation}
and thus by the chain rule:
\begin{equation}
    \begin{aligned}
        \frac{\partial \mathcal{V}(x^{eq})}{\partial \lambda} &= 
        f'(g(\lambda)) g'(\lambda) = 
        ( - \overline{d} - 2  \frac{(\lan)^2}{(1+\lan)^2} d') 
        \cdot \frac{2 \lambda n^2}{(1 + \lan)^3} \\
        &= \frac{2 \lambda n^2}{(1 + \lan)^5} \biggl ( 
            - \overline{d} (1 + 2 \lan + \lambda^2 n^2) - 2 \lambda^2 n^2 d'
        \biggr ) \\
        &= \underbrace{\frac{2 \lambda n^2}{(1 + \lan)^5}}_{\displaystyle \coloneqq C(\lambda)} 
        \underbrace{\biggl (
            - n^2 (\overline{d} + 2 d') \lambda^2 - 2n \overline{d} \lambda - \overline{d}
        \biggr )}_{\displaystyle \coloneqq h(\lambda)}
    \end{aligned}
\end{equation}

We now aim to determine the range of $\lambda$ values for which the derivative is positive and the range for which it is negative. As $C(\lambda) > 0$ for all $\lambda > 0$ (recall we are only interested in positive $\lambda$ values), the sign of the derivative is simply the sign of $h(\lambda)$.
We define the condition $\overline{d} + 2d' \geq 0$ as the regularity condition and divide our analysis into two cases:

\begin{itemize}
    \item \textbf{Case I - the regularity condition holds ($\overline{d} + 2d' \geq 0$)}: In this case, $h(\lambda) \leq 0 \; \forall \lambda > 0$. So $\frac{\partial \mathcal{V}(x^{eq})}{\partial \lambda} \leq 0 \; \forall \lambda > 0$ and therefore in this case the users' welfare in equilibrium is monotonically decreasing in $\lambda$.

    \item \textbf{Case II - the regularity condition does not hold ($\overline{d} + 2d' < 0$)}: In this case $h(\lambda)$ is a strictly convex quadratic function, so it suffices for us to find its roots in order to know its positivity and negativity ranges. By the quadratic formula, $h$'s roots are:

    \begin{equation}
    \begin{aligned}
        \lambda_{1,2} &= \frac{2n \overline{d} \pm
        \sqrt{(- 2n \overline{d})^2 - 4 \bigr(- n^2 (\overline{d} + 2 d') \bigl) (- \overline{d})}}
        { - 2 n^2 (\overline{d} + 2 d')}
        \\ &= \frac{2n \overline{d} \pm
        \sqrt{- 8 n^2 \overline{d} d'}}
        { - 2 n^2 (\overline{d} + 2 d')} = \frac{\overline{d} \pm
        \sqrt{- 2 \overline{d} d'}}
        { - n (\overline{d} + 2 d')}
    \end{aligned}
    \end{equation}

    Let us show that the term inside the square root is strictly positive. First, $\overline{d} \geq 0$ as the average of non-negative terms. $\overline{d} = 0$ would imply that $d^i=0 \; \forall i$ and then we get $d' = 0$ which contradicts $\overline{d} + 2d' < 0$. Therefore, $\overline{d} > 0$. Additionally, from $\overline{d} + 2d' < 0$ we get 
    $\overline{d} < - 2d'$ and thus also $0<- 2d'$. So $- 2 \overline{d} d' > 0$. Now we know that both $\lambda_1$ and $\lambda_2$ are well-defined and that $h$ has two distinct roots.

    Furthermore, we notice that 
    \begin{equation}
        \lambda_1 \coloneqq \frac{\overline{d} +
        \sqrt{- 2 \overline{d} d'}}
        { - n (\overline{d} + 2 d')} > 0
    \end{equation} 
    since the numerator is strictly positive ($\overline{d} > 0$) and the denominator is strictly positive ($\overline{d} + 2d' < 0$). 

    As for the other root, 
    \begin{equation}
        \lambda_2 \coloneqq \frac{\overline{d} -
        \sqrt{- 2 \overline{d} d'}}
        { - n (\overline{d} + 2 d')} < 0
    \end{equation}
    because $\lambda_2 < 0 \iff
    \overline{d} - \sqrt{- 2 \overline{d} d'} < 0 \iff 
    \overline{d} < \sqrt{- 2 \overline{d} d'} \underset{(1)}{\iff}
    \overline{d}^2 < - 2 \overline{d} d' \iff
    \overline{d}^2 + 2 \overline{d} d' < 0 \underset{(2)}{\iff} 
    \overline{d} + 2d' < 0$ where both (1) and (2) can be justified by $\overline{d} > 0$, which we already showed.

    Summing up this case, we got that $h(\lambda)$ is a strictly convex quadratic function with roots $\lambda_2 < 0 < \lambda_1$. So $h(\lambda)$ is negative when $0<\lambda<\lambda_1$ and positive when $\lambda_1 < \lambda$. Therefore, these are also the negativity and positivity ranges of $\frac{\partial \mathcal{V}(x^{eq})}{\partial \lambda}$. Thus, $\lambda_1$ is a global minimum point of $\mathcal{V}(x^{eq})$.
    
\end{itemize}

\begin{remark}
    The regularity condition \(\overline{d} + 2d' \geq 0\) is equivalent to requiring that the unbiased sample variance of \(\{d^i\}_{i \in N}\) is no greater than half the empirical mean \(\overline{d}\). This follows from the fact that $d' = \frac{n}{n-1} \left( \overline{d}^2 - \frac{1}{n} \sum_{i=1}^n (d^i)^2 \right)$,
    which shows that \(d'\) is equal to minus the unbiased sample variance of \(\{d^i\}_{i \in N}\).
\end{remark}

\begin{figure}
\centering
\includegraphics[width=.4\textwidth,height=\textheight,keepaspectratio]{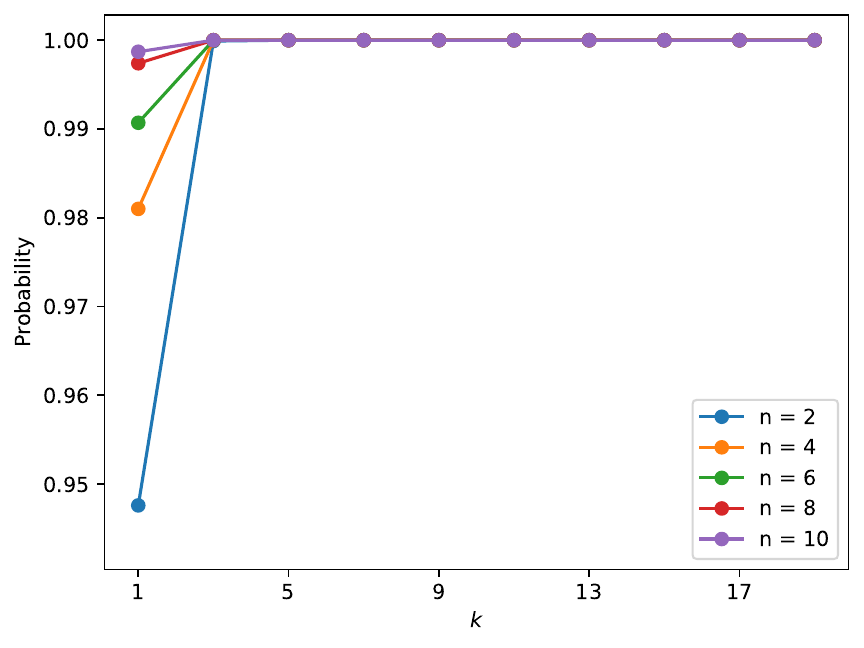}
\caption{Probability of regularity condition with independently and uniformly distributed $x_0$ and $x^*$.}
\label{fig: reg_prob}
\Description[Probability of the regularity condition for different values of k and n]{Importantly, the probability is always above 0.94. Both an increase in n and an increase in k cause the probability of the condition to increase.}
\end{figure}

In Figure \ref{fig: reg_prob} we estimate the probability for the regularity condition for different $n$ and $k$ values with independently and uniformly distributed $x_0$ and $x^*$.

To sum up, in this section we have demonstrated how the closed form of the equilibrium allows us to easily analyze the effect of $\lambda$ on the welfare measures in the PNE. A similar analysis can be performed on the effect of the other game parameters.

\section{The Effect of the Embedding Space Dimension}\label{ape: k effect}

In this section, we discuss the effect of the embedding space dimension, $k$. This examination is more complicated than examining the effect of $\lambda$ since the time complexity of the simulation increases exponentially with $k$. We first discuss the effect of $k$ when using the softmax and the linear ranking functions we explored in \S\ref{sec: empirical} and then present two approaches to assess this effect when using the PRP ranking function, a direct approach and a dynamics-based approach. Finally, we discuss how the results of this section affect the comparison between the different ranking functions. Note that $k=1$ is essentially different than other $k$ values since it is the only case in which the possible movement directions in the Discrete Better Response Dynamics Simulation include all possible directions in the embedding space. This attribute, together with the fact that $k=1$ is not a realistic case regardless, leads us to omit it from our analysis.

\subsection{Direct Examination of the Effect of $k$ on the Softmax and the Linear Ranking Functions}

\begin{figure}
\centering
\includegraphics[width=\textwidth,height=\textheight,keepaspectratio]{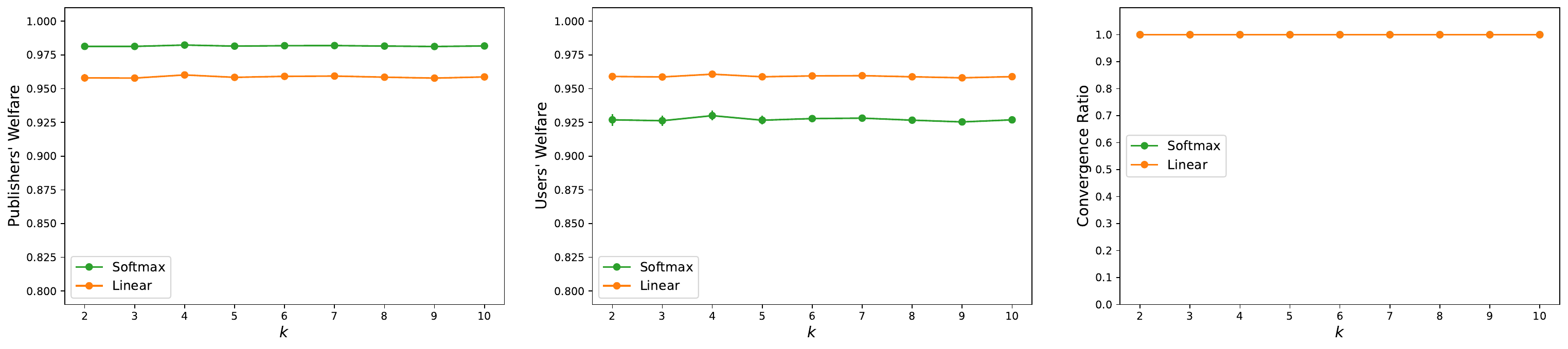}
\caption{The effect of $k$ in the softmax ($\beta = 1$) and the linear ($a = \frac{1}{n} = 0.5$) ranking functions with $\lambda = 0.5$ and $ n = 2$.}
\label{fig: k_rrp}
\Description[Figure 6]{Completely described in the text.}
\end{figure}

We performed $500$ simulations for each \(k\) value to directly test the effect of $k$ on the welfare measures and on the convergence ratio under the softmax and the linear ranking functions.\footnote{In these simulations we used the same simulations' parameters as in the simulations presented in \S\ref{sec: empirical}. \\
$\mathcal{S} = \{2^{-6}, 2^{-5}, ..., 2^{-2}, 0.5, 0.6, ..., 1 \}, $ $\mathcal{D} = \big \{ \frac{d}{||d||_2}: d \in \{-1,0,1\}^k \setminus \{ \vec{0} \} \big \}, \; T=1000, \; M=900$ and $\varepsilon=10^{-6}$, with a bootstrap sample size of $B = 500$.} The results of these simulations are presented in Figure \ref{fig: k_rrp}.

We can see that the convergence ratios of both ranking functions remain $1$ for all tested $k$ values. Furthermore, Figure \ref{fig: k_rrp} evidences that when using the softmax or the linear ranking functions $k$ has a negligible effect on both the publishers' and the users' welfare. We observed similar results for other $\lambda$ values and different ranking functions hyperparameters. Therefore, it can be said that the results in \S\ref{sec: empirical} regarding those functions can be extrapolated to any low-mid $k$ value. Examining high values of $k$ is highly impractical using this kind of simulation due to the exponential growth of the simulation running time as $k$ increases. However, note that all our theoretical results regarding the linear ranking function hold for any $k$. In particular, the convergence of any $\varepsilon$-better response dynamics to a $\varepsilon$-PNE is guaranteed, and thus the convergence ratio of the linear ranking function is guaranteed to be $1$ for any $k$.

Furthermore, in Appendix \ref{ape: euclidean norm} we presented Lemma \ref{lemma: the line dominates}, which determines that in games induced by various ranking functions, including the softmax and the linear ranking functions, publishers should never play a strategy (document) that is not on the line segment that connects $x^*$ with their initial document. This result shows us that the nature of the dynamics in publishers' games induced by these ranking functions is highly one-dimensional. That is, no matter how large $k$ is, each player still plays roughly on her one-dimensional line segment. Thus, we conjecture that the softmax ranking function will have a convergence ratio of $1$ for all $k$. 
Furthermore, based on this argument, we can also conjecture that $k$ will have a negligible effect on the welfare measures of both the linear and softmax ranking functions.

\subsection{Direct Examination of the Effect of $k$ on the PRP Ranking Function}

\begin{figure}[tb]
\centering
\includegraphics[width=\textwidth,height=\textheight,keepaspectratio]{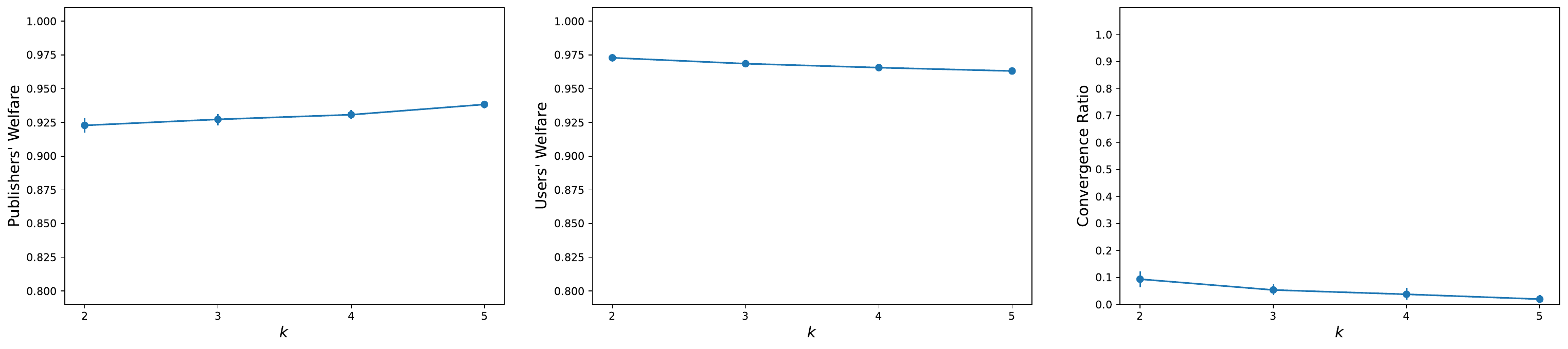}
\caption{The effect of k in the PRP ranking function with $\lambda = 0.5$ and $ n = 2$.}
\label{fig: k_prp}
\Description[Figure 7]{Completely described in the text.}
\end{figure}

We now turn to discuss the effect of $k$ in PRP-induced publishers' games, which introduces an increase in analytical complexity. In the Discrete Better Response Dynamics Simulation, the number of movement directions, and hence the simulation running time, grows exponentially with $k$. This challenge is further amplified when simulating the PRP ranking function, as the simulations frequently fail to converge. This makes simulations of the PRP ranking function with high values of $k$ highly impractical. Figure \ref{fig: k_prp} presents the results for a few low values of $k$.\footnote{In this examination we ran 500 simulations for each \(k\) value, with $\mathcal{S} = \{2^{-6}, 2^{-5}, ..., 2^{-2}, 0.5, 0.6, ..., 1 \},$ \\ $\mathcal{D} = \big \{ \frac{d}{||d||_2}: d \in \{-1,0,1\}^k \setminus \{ \vec{0} \} \big \}, \; T=5000, \; M=4500$ and $\varepsilon=10^{-6}$, with a bootstrap sample size of $B = 500$. We chose high $T$ and $M$ values since the cycle length of the pseudo-periodic dynamics of the PRP ranking function (which we discussed in \S\ref{subsec: learning dynamics}) increases as $k$ increases.}

Figure \ref{fig: k_prp} may indicate that, as is the case for the softmax and the linear ranking functions, $k$ has a minor effect on the publishers' and users' welfare when using the PRP ranking function. We can see that as $k$ increases, the publishers' welfare slightly increases and the users' welfare slightly decreases. As mentioned, due to the structure of the simulation, we can't examine the welfare measures with higher $k$ values, but we conjecture that the observed trends will continue until a certain limit.

Regarding the convergence ratio, Figure \ref{fig: k_prp} might suggest that it generally decreases with $k$ and that the decrease becomes more moderate as $k$ increases. As we only tested low values of $k$, it remains uncertain whether this trend holds for higher values of $k$ as well. Therefore, we employ an alternative approach to approximate the effect of $k$ on the convergence ratio of the PRP ranking function.

\subsection{Dynamics-Based Examination of the Effect of $k$ on the Convergence Ratio of the PRP Ranking Function}

\begin{figure}
\centering
\includegraphics[width=.3\textwidth,height=\textheight,keepaspectratio]{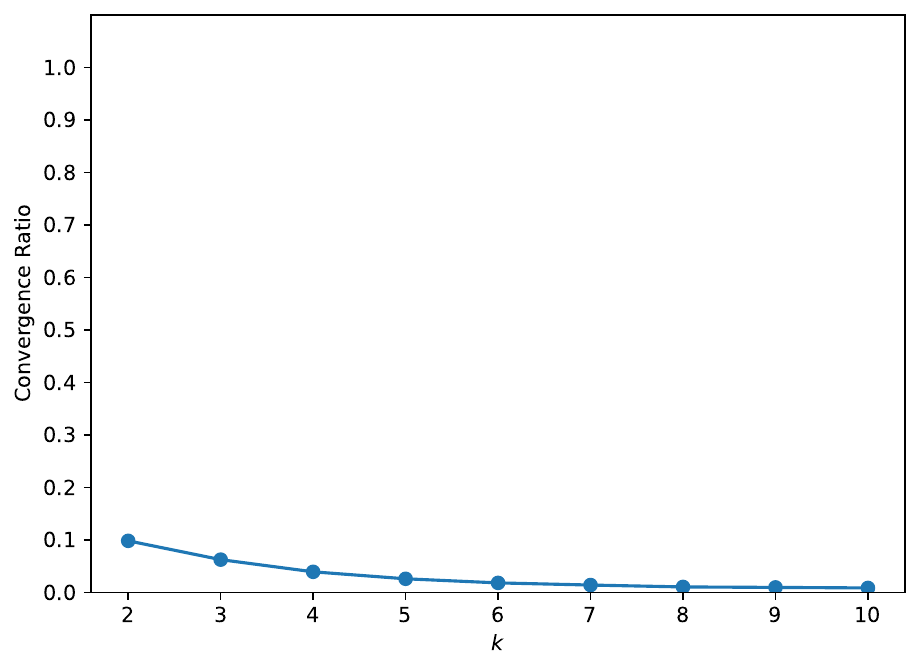}
\caption{The effect of k on the PRP ranking function's convergence ratio with $\lambda = 0.5$ and $ n = 2$.}
\label{fig: k_prp_converge}
\Description[Figure 8]{The convergence ratio starts at approximately 0.1 and decreases as k increases.}
\end{figure}

In \S\ref{subsec: learning dynamics} we observed the two primary types of dynamics induced by the PRP ranking function, the uncommon type where the simulation converges already in the first timestep, and the common type where the publishers reach a pseudo-periodic behavior. Based on these results, we can use a new method to approximate the effect of $k$ on the convergence ratio of the PRP ranking function. Instead of running a whole simulation each time, we can determine convergence at the beginning of the simulation. If the dynamics did not converge in the first timestep, we can deduce that the simulation will not converge. Figure \ref{fig: k_prp_converge} presents the convergence ratio measured using this estimation method.\footnote{In this special method we have conducted \(10,000\) simulations with $\mathcal{S} = \{2^{-6}, 2^{-5}, ..., 2^{-2}, 0.5, 0.6, ..., 1 \},$ \\ $\mathcal{D} = \big \{ \frac{d}{||d||_2}: d \in \{-1,0,1\}^k \setminus \{ \vec{0} \} \big \}$ and $\varepsilon=10^{-6}$, and used a bootstrap sample size of $B = 10,000$. We have conducted a large number of simulations, since the measurement of the convergence ratio involves a lot of noise, and we could not use the same group of randomized $x_0$ and $x^*$ for all the $k$ values due to the different vector dimensions. Conducting a large number of simulations was made possible thanks to our method of running each simulation for a single timestep.}

Figure \ref{fig: k_prp_converge} shows that the convergence ratio of the PRP ranking function generally decreases as $k$ increases and that the decrease becomes more moderate as $k$ increases. In \S\ref{subsec: learning dynamics}, we saw that the case when PRP dynamics converge is when the publisher whose initial document is farther from $x^*$ cannot bypass the other publisher in a single step. Therefore, a possible explanation for the decrease in the convergence ratio is that a larger $k$ decreases the likelihood of this case. Although Figure \ref{fig: k_prp_converge} provides results only for $\lambda=0.5$, similar trends were observed for other values of $\lambda$. We conjecture that these trends continue at higher $k$ values.

To conclude this section, let us revisit the trends we discovered in light of the results we have already seen in \S\ref{sec: empirical}. From this perspective, we can see how an increase in $k$ emphasizes the consequences of the instability of the PRP ranking function. The already low convergence ratio at $k=2$ further diminishes with increasing $k$. Simultaneously, the users' welfare, a measure that the PRP is designed to maximize in the short term, also experiences a minor decline as $k$ increases. In contrast, the examination of different $k$ values underscores the stability of both the softmax and the linear ranking functions, maintaining a convergence ratio of $1$ and roughly identical welfare measures across all tested $k$ values.

\section{Additional Experiments}
In this section, we provide and discuss additional experimental results. These experiments aim to provide some preliminary insights regarding settings beyond the ones presented in the body of the paper. In all simulations in this section, we used the same simulation parameters as in the simulations presented in \S\ref{sec: empirical}.

\subsection{Beyond the Two-Publisher Case}\label{app: beyond n=2}

\begin{figure}
\centering
\includegraphics[width=\textwidth,height=\textheight,keepaspectratio]{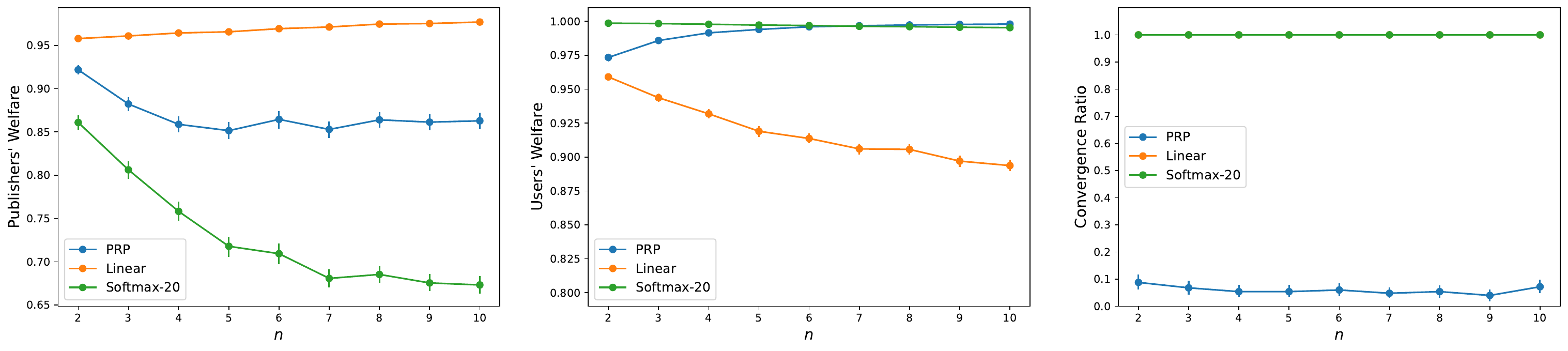}
\caption{Comparison of the PRP, softmax ($\beta = 20$) and linear ($a=\frac{1}{n}$) ranking functions across different $n$ values with constant $\lambda = \frac{1}{2}$.
The linear and the softmax ranking functions had a constant convergence ratio of $1$.}
\label{fig: n_comp}
\Description[Figure 9]{Completely described in the text.}
\end{figure}

Figure \ref{fig: n_comp} depicts the effect of $n$ on our three evaluation criteria (publishers’ welfare, users’ welfare, and convergence ratio), under the three ranking schemes.
Importantly, the slope of the linear ranking function decreases as $n$ grows large (that is, $a=\frac{1}{n}$). As a result, the performance of the linear ranking function deteriorates in terms of users’ welfare as the number of publishers grows. Note that for lower values of $n$, the softmax ranking function (with inverse temperature $\beta = 20$) outperforms the PRP ranking in terms of users’ welfare.
However, as $n$ grows, the users' welfare under PRP ranking improves until it reaches the users' welfare under softmax ranking. That being said, it might be the case that adaptively choosing $\beta$ as a function of $n$ can yield a consistent improvement over the PRP in terms of users' welfare.
The stability of the linear and softmax schemes (as well as the instability of the PRP) is also preserved.

The decrease in users’ welfare under linear ranking is expected, since the slope $a=\frac{1}{n}$ decreases as $n$ increases, limiting the extent to which the ranker can incentivize publishers towards relevant content (see the discussion in \S\ref{linear_theory}). Having said that, we highlight that the case of small $n$ (even $n=2$) is widely observed when publishers compete in a niche domain. This also aligns with the assumption of a single, known $x^*$ – publishers specialize in a specific domain and know precisely the users’ preferences and required content. 
For higher values of $n$, the softmax ranking function (with an appropriate $\beta$) provides a better alternative in terms of users' welfare since $\beta$ can be taken arbitrarily large.

\subsection{Beyond the Uniform i.i.d. Instance Distribution Case}\label{app: non iid dist}

We now consider a different family of instance distributions, on which we evaluate the three ranking schemes. These distributions diverge from the assumptions of independence between the initial documents and the information need, as well as the assumption of uniform marginal distributions.
Specifically, we sample from a multivariate normal distribution, conditioned to lie within $[0,1]^k$.
For each coordinate $j \in [k] = [2]$, the $j$-th entries of the $n=2$ publishers’ initial documents and the information need are sampled from the conditional multivariate normal distribution with mean $(0.5, 0.5, 0.5)$, and covariance matrix of:

\[
\begin{array}{c|ccc}
   & \text{[x}_0^1]_j & \text{[x}_0^2]_j & \text{[x}^*]_j \\
\hline
\text{[x}_0^1]_j & 1 & \rho_1 & \rho_2 \\
\text{[x}_0^2]_j & \rho_1 & 1 & \rho_1 \rho_2 \\
\text{[x}^*]_j & \rho_2 & \rho_1 \rho_2 & 1 \\
\end{array}
\]

\begin{figure}[H]
    \centering

    \includegraphics[width=\textwidth]{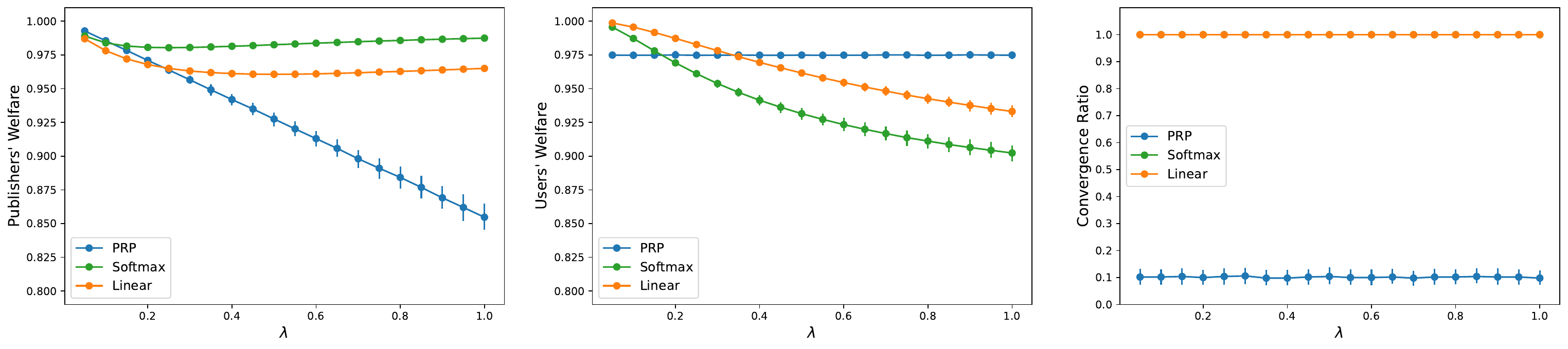}
    \caption*{(a) $\rho_1 = \rho_2 = 0$}

    \includegraphics[width=\textwidth]{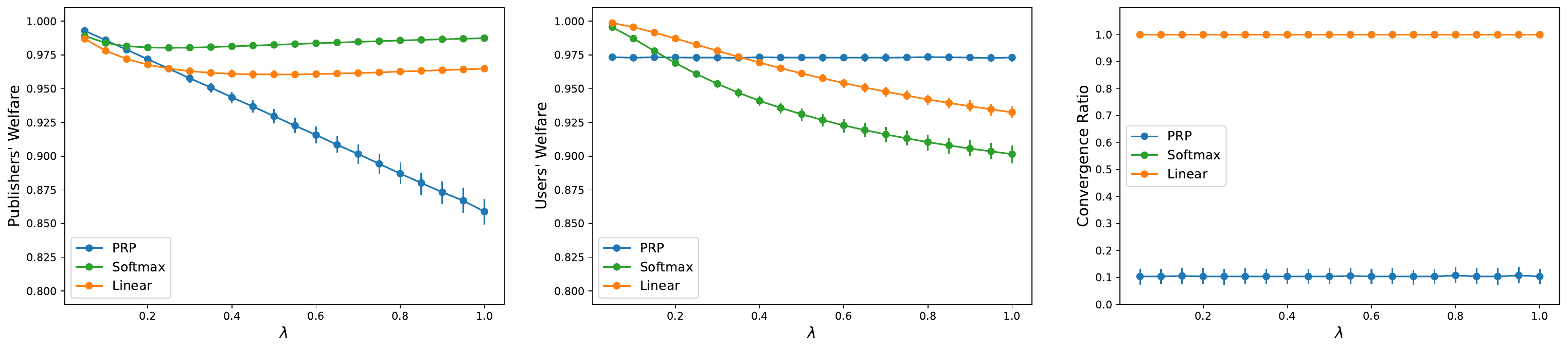}
    \caption*{(b) $\rho_1 = \frac{1}{2}, \; \rho_2 = 0$}\label{fig: normal_comp_ro1}

    \includegraphics[width=\textwidth]{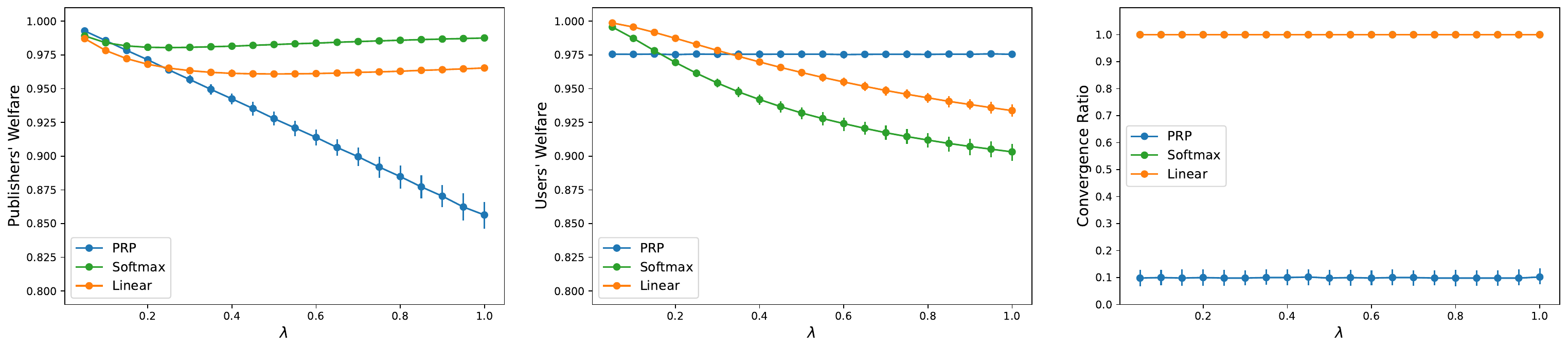}
    \caption*{(c) $\rho_1 = 0, \; \rho_2 = \frac{1}{2}$}\label{fig: normal_comp_ro2}

    \includegraphics[width=\textwidth]{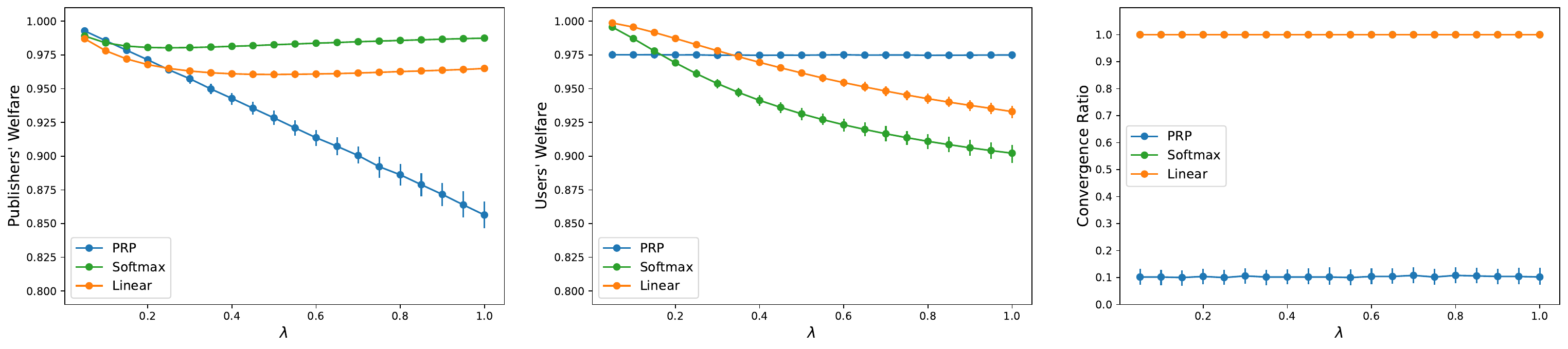}
    \caption*{(d) $\rho_1 = \frac{1}{2}, \; \rho_2 = \frac{1}{2}$}\label{fig: normal_comp_ro1_ro2}

    \caption{Comparison of the PRP, softmax ($\beta = 1$) and linear ($a=\frac{1}{n}$) ranking functions with initial documents and the information need being distributed according to the conditional multivariate normal distribution. The parameters of the distribution are provided below each figure.
    In all cases, The linear and the softmax ranking functions had a constant convergence ratio of $1$.}
    \Description[Figure 10]{Completely described in the text.}
    \label{fig: distribution_plots}

\end{figure}

Note that while entries in different coordinates remain independent, there is a correlation between the initial documents and the information need within each coordinate.
When $\rho_1=\rho_2=0$, this distribution corresponds to a conditional multivariate normal distribution with the initial documents and the information need being pairwise independent. In Figure \ref{fig: distribution_plots}a we present the trends of publishers’ welfare, users’ welfare, and convergence ratio as a function of $\lambda$ for $\rho_1=\rho_2=0$. Note that the trends are as in the uniform case presented in \S\ref{subsec: simulation results - between rankers}.

Next, we introduce three additional cases, in which there is a correlation between the initial documents and the information need. The first case, presented in Figure \ref{fig: distribution_plots}b, corresponds to $\rho_1=\frac{1}{2}, \rho_2=0$. Intuitively, this means that the publishers' initial documents are positively correlated. 
The second case, presented in Figure \ref{fig: distribution_plots}c, corresponds to $\rho_1=0, \rho_2=\frac{1}{2}$, which means that the initial document of publisher $1$ will be generally closer to the information need. 
The third case, presented in Figure \ref{fig: distribution_plots}d, corresponds to $\rho_1=\frac{1}{2}, \rho_2=\frac{1}{2}$, which means that both trends of the first two cases occur, and as a result there is some weaker correlation between the initial document of publisher $2$ and the information need.

Notice that in all three cases, the trends again agree with the ones presented in the paper.
While these results are preliminary, we believe similar trends are preserved in many other natural instance distributions as well.

\newpage

\section{Omitted Proofs of the Auxiliary Lemmas}
\label{ape: proofs}

\begin{proof}[Proof of Lemma \ref{lemma: prp trivial pne}]
    Assume by contradiction that there exists $i \in N$ such that $x^{eq}_i \notin \{ x^*, x_0^i\}$. 
    By Lemma \ref{lemma: the line dominates} we get that $x^{eq}_i \in L_i \coloneqq \{t x^* + (1 - t)x_0^i : t \in [0,1]\}$. 
    Thus $x_0^i \neq x^*$. \\
    Recall the notation $\mu(x) \coloneqq \argmin_{i \in N}{d^*(x_i)}$ for $x \in X$.
    Let us divide the proof into the following cases:
    \begin{enumerate}
        \item If $i \notin \mu(x^{eq})$ then $u_i(x^{eq}) = -\lambda d^0_i(x^{eq}_i)$.
        Since $x^{eq}_i \neq x_0^i$ we get that $d^0_i(x^{eq}_i) > 0$ and hence $u_i(x^{eq}) < 0$.
        On the other hand, $u_i(x_0^i, x^{eq}_{-i}) = r_i(x_0^i, x^{eq}_{-i}) - 0 \geq 0$.
        Hence $u_i(x_0^i, x^{eq}_{-i}) > u_i(x^{eq})$ which is a contradiction to the fact that $x^{eq}$ is a PNE.
        \item If $\mu(x^{eq}) = \{i\}$ then $d^*(x^{eq}_i) < d^*(x^{eq}_j) \; \forall j \neq i$.
        Let us denote $\varepsilon = \frac{1}{2} \bigl ( \min_{j \neq i}{d^*(x^{eq}_j)} - d^*(x^{eq}_i) \bigr ) > 0$.
        Since $d$ is continuous, $d^*$ is continuous and hence there exists a $\delta > 0$ 
        such that if $\norm{x_i - x^{eq}_i}_2 < \delta$ then $d^*(x_i) < d^*(x^{eq}_i) + \varepsilon$.
        Therefore, it holds in particular for 
        \begin{equation}
                        \hat{x}_i \coloneqq x^{eq}_i + \frac{\delta}{2 \norm {x^{eq}_i - x_0^i}_2} (x_0^i - x^{eq}_i)
        \end{equation}

        that $d^*(\hat{x}_i) < d^*(x^{eq}_i) + \varepsilon$ and hence 
        $d^*(\hat{x}_i) < \min_{j \neq i}{d^*(x^{eq}_j)}$.
        Therefore $\mu(\hat{x}_i, x^{eq}_{-i}) = \{i\}$ and hence
        $r^*_i(\hat{x}_i, x^{eq}_{-i}) = 1$. We get that:
        \begin{equation}
        \begin{aligned} 
        u_i(\hat{x}_i, x^{eq}_{-i}) &= 1 - \lambda d^0_i(\hat{x}_i) =
        1 - \lambda \frac{1}{k} \norm{\hat{x}_i - x_0^i}_2^2 =
        1 - \lambda \frac{1}{k} \norm{(x^{eq}_i - x_0^i) (1-\frac{\delta}{2 \norm {x^{eq}_i - x_0^i}_2})}_2^2 \\
        &=
        1 - \lambda \frac{1}{k}(1-\frac{\delta}{2 \norm {x^{eq}_i - x_0^i}_2})^2 \norm{x^{eq}_i - x_0^i}_2^2
        >
        1 - \lambda \frac{1}{k} \norm{x^{eq}_i - x_0^i}_2^2 \\
        &= 1 - \lambda d^0_i(x^{eq}_i)= u_i(x^{eq})
        \end{aligned}
        \end{equation}
        
        which is in contradiction to the fact that $x^{eq}$ is a PNE.

        \item If $i \in \mu(x^{eq})$ and $|\mu(x^{eq})| \geq 2$: \\
        $d$ is continuous and hence $d^0_i$ is continuous. Therefore there exists a $\delta > 0$ 
        such that if $\norm{x_i - x^{eq}_i}_2 < \delta$ then $d^0_i(x_i) < d^0_i(x^{eq}_i) + \frac{1}{2 \lambda}$.
        Therefore, it holds in particular for
        \begin{equation}
            \hat{x}_i \coloneqq x^{eq}_i + \frac{\delta}{2 \norm {x^{eq}_i - x^*}_2} (x^* - x^{eq}_i)
        \end{equation}

        that $d^0_i(\hat{x}_i) < d^0_i(x^{eq}_i) + \frac{1}{2 \lambda}$.

        Now, $d^*(\hat{x_i}) = \frac{1}{k} \norm{\hat{x}_i - x^*}_2^2 =
        \frac{1}{k} \norm{(x^{eq}_i - x^*)(1 - \frac{\delta}{2 \norm {x^{eq}_i - x^*}_2})}_2^2 < \frac{1}{k} \norm{x^{eq}_i - x^*}_2^2 = d^*(x^{eq}_i)$.
        Therefore $\mu(\hat{x}_i, x^{eq}_{-i}) = \{i\}$ and hence $r_i(\hat{x}_i, x^{eq}_{-i}) = 1$.
        We get that:
        \begin{equation}  
        \begin{aligned} 
        u_i(\hat{x}_i, x^{eq}_{-i}) & = 1 - \lambda d^0_i(\hat{x}_i) >
            1 - \lambda \bigr (d^0_i(x^{eq}_i) + \frac{1}{2 \lambda} \bigl) 
            \\ & =
            \frac{1}{2} - \lambda d^0_i(x^{eq}_i) \geq
            \frac{1}{|\mu(x^{eq})|} - \lambda d^0_i(x^{eq}_i) =
            u_i(x^{eq})
            \end{aligned}
        \end{equation}
        which is once again a contradiction to the fact that $x^{eq}$ is a PNE.
    \end{enumerate}
\end{proof}

\begin{proof}[Proof of Lemma \ref{lemma: second order condition}]
    If $g$ is constant on $[0,1]$ then $f_i \equiv \frac{1}{n} \; \forall i \in [n]$ and then $f$ satisfies \eqref{double derivatives} trivially, since both sides of the equation are $0$.
    \\In the other direction, assume that $f$ satisfies \eqref{double derivatives}. Let us calculate the derivatives. It holds $\forall i,j\in [n]$ and $\forall x \in X$ that:
    \begin{equation}
        \frac{\partial f_i (x)}{\partial x_j} = 
    \frac{-g(x_i) g'(x_j)}{\left(\sum\limits_{k=1}^n g\left(x_k\right)\right)^2}
    \end{equation} 
    and
    \begin{equation}
        \frac{\partial^2 f_i (x)}{\partial x_i \partial x_j} = 
    \frac{
      g'(x_i)g'(x_j)\left( 2g(x_i) - \sum\limits_{k=1}^n g(x_k) \right)
    }{
      \left(\sum\limits_{k=1}^n g(x_k)\right)^3
    }
    \end{equation}

    In the same way, we get:
    \begin{equation}
        \frac{\partial^2 f_j (x)}{\partial x_i \partial x_j} =
    \frac{
      g'(x_i)g'(x_j)\left( 2g(x_j) - \sum\limits_{k=1}^n g(x_k) \right)
    }{
      \left(\sum\limits_{k=1}^n g(x_k)\right)^3
    }
    \end{equation}
    
    So by \eqref{double derivatives} we get:
    \begin{equation}
         0 = \frac{\partial^2 f_i(x)}{\partial x_i \partial x_j} -
     \frac{\partial^2 f_j(x)}{\partial x_i \partial x_j}
    = \frac{
        g'(x_i)g'(x_j)\left( 2g(x_i) - 2g(x_j) \right)
    }{
        \left(\sum\limits_{k=1}^n g(x_k)\right)^3
    } 
    \end{equation}

    Hence, $\forall i,j\in [n]$ and $\forall x \in [0,1]^n$ it holds that:
    \begin{equation}
        g'(x_i) = 0 \; \vee \; g'(x_j) = 0 \; \vee \; g(x_i) = g(x_j) 
    \end{equation}

    Equivalently, (since $n \geq 2$):
    \begin{equation} \label{constant} g'(a) = 0 \; \vee \; g'(b) = 0 \;
      \vee \; g(a) = g(b)    \;\;\; \forall a,b\in [0,1]
    \end{equation}

    We proceed to show that \eqref{constant} implies that $g$ is constant on $[0,1]$.
    We divide the proof into two cases:
    \begin{itemize}
      \item
    If there is no $a \in [0,1]$ such that
    $g'(a) \neq 0$, then $g' \equiv 0$ on $[0,1]$ 
    and therefore $g$ is constant on $[0,1]$.
    \item
    Otherwise, let $a \in [0,1]$ be such that $g'(a) \neq 0$.
    Without loss of generality, assume $g'(a) > 0$.
    By continuity of $g'$, there is a $\delta > 0$ such that
    $g'(x) > 0 \; \forall x \in U:=(a-\delta, a+\delta)\cap [0,1]$.
    So $g$ is strictly increasing on $U$.
    Take $b \in U$.
    Since $g$ is strictly increasing on $U$,
    $g(a) \neq g(b)$. Also, $g'(b) \neq 0$, $g'(a) \neq 0$,
    since they are both positive. This is a contradiction to \eqref{constant}.

    \end{itemize}

\end{proof}    

\begin{proof}[Proof of Lemma \ref{lemma: psi property}]
We have that:
    \begin{equation}
        \psi_i(d) = \frac{1}{n-1} \sum\limits_{j\neq i} d_j - d_i
    = \frac{1}{n-1} \sum\limits_{j=1}^n d_j - \frac{n}{n-1}  d_i 
    \end{equation}
Therefore:
    \begin{equation}
        \psi_i(d) = \psi_j(d) \iff \frac{n}{n-1}d_i = \frac{n}{n-1}d_j \iff d_i = d_j
    \end{equation}
\end{proof}


\vspace{1em}

\end{document}